\documentclass[final]{IEEEtran}
\usepackage{color}
\usepackage{array}
\usepackage{cite}
\usepackage{amssymb}
\usepackage{amsmath}
\usepackage{amsthm}
\usepackage[dvips]{graphicx}
\usepackage[font=small]{caption}
\usepackage{psfrag}
\usepackage{url}
\usepackage{subfigure}
\usepackage{slashbox}

\newtheorem{definition}{Definition}

\newtheorem{proposition}{Proposition}
\newcommand{\dprime}{{\prime\prime}}
\newcommand{\new}[1]{{\color{black}#1}}
\newcommand{\rev}[1]{{\color{black}#1}}

\newcommand{\edt}[1]{{\color{black}#1}}
\newcommand{\rvn}[1]{{\color{black}#1}}
\title{
Generalized Cross-Layer Designs for Generic Half-Duplex Multicarrier Wireless Networks with Frequency-Reuse
}
\author{Rozita Rashtchi, \IEEEmembership{Student Member, IEEE}, Ramy H. Gohary, \IEEEmembership{Senior Member, IEEE}, and Halim  Yanikomeroglu, \IEEEmembership{Senior Member, IEEE}
\thanks{The authors are with the Department of Systems and Computer Engineering, Carleton University,  Ottawa, Ontario, Canada,
\{rozita, gohary, halim\}@sce.carleton.ca}
\thanks{Preliminary versions of this work were presented, in part, at the \emph{IEEE Glob. Telecommun.~(Globecom)}, 2013, and, in part, at the \emph{IEEE Int. Wkshp. Signal Processing Advances in Wireless Commun.~(SPAWC)}, 2014.}
\thanks{This work is supported in part by {a}  Discovery Grant~(DG)  of the Natural Sciences and Engineering Research Council~(NSERC) of Canada, in part  by Huawei Canada Co., Ltd., and in part by the Ontario Ministry of Economic Development and Innovation’s ORF-RE
(Ontario Research Fund - Research Excellence) program.}
}
\allowdisplaybreaks
\begin{document}
\maketitle
\begin{abstract}
In this paper, joint designs of data routes and resource allocations are developed for  generic half-duplex multicarrier wireless networks  in which each subcarrier can be reused by multiple links. 
Two instances are considered. \rvn{The first instance pertains to the general case in which each subcarrier can be time-shared by multiple links,
whereas the second instance pertains to a special case in which time-sharing is not allowed and a subcarrier, once assigned to a set of links, is used by those  links throughout the signalling interval.
}
Novel frameworks are developed to optimize the joint design of data routes, subcarrier schedules and power allocations. 
These design problems are nonconvex and hence difficult to solve. To circumvent this difficulty, efficient  techniques based on geometric programming are developed to obtain  locally optimal solutions.
%
%
Numerical results show that the designs developed in both instances yield performance that is superior to that of their counterparts in which frequency-reuse is not allowed.
\end{abstract}
\section{Introduction} \label{se:intro}
The prospect of having ubiquitous high data-rate  wireless services is leveraged by the versatility and portability of the communication devices that will form the nodes of  future wireless networks. These devices will  be able to perform various functions including sending, receiving and/or relaying data to other nodes. As such, it is expected that future wireless networks will not possess a predetermined topology, but rather \rvn{an \textit{ad hoc}}
one that encompasses many existing and upcoming network structures 
 including current and relay-aided cellular networks~\cite{Halim02,BHW03}.

Given the stringent limitations on the spectrum available for wireless communications, providing high data-rate services banks on sharing the spectrum by multiple users which results in
 potentially significant interference. 
One option to mitigate interference is to use the
 Orthogonal Frequency Division Multiple Access~(OFDMA) technique, wherein
 a set of  orthogonal narrow-band subcarriers are exclusively assigned to each user.
 This technique offers
several 
advantages
 including
 design simplicity and resilience to frequency-selective fading. 
In spite of these advantages,  \rvn{rate-effective} utilization of the available spectrum may require the OFDMA subcarriers to be used simultaneously, rather than exclusively,  by multiple users.
This is especially the case when the network is composed of essentially separated clusters.  
\rvn{In contrast, for tightly coupled networks exclusive usage of subcarriers can be  more beneficial from a rate perspective~\cite{HL09},~\cite{LL07}}.

In order for  a wireless network to be able to support the reliable communication of high data rates, the scarce resources available for the network must be carefully exploited. Such resources include the spectrum available for communication, time, and the typically low power of the wireless nodes. Proper exploitation of these resources involves choosing the optimal routes of the  data flows, the optimal  powers to be allocated by the nodes to each subcarrier, and the optimal scheduling  and possibly duration  over which the subcarriers are assigned to various links. 
Although  these tasks have traditionally been performed separately, they  are interrelated and performing them in isolation
may incur a significant loss in performance. 

To avoid the aforementioned drawback,
 we devise  a joint optimization framework that incorporates data routing, subcarrier scheduling and power allocation in the design of a generic  multicarrier network. \rvn{The network is generic in the sense that it possesses an \textit{ad hoc} topology and its nodes can assume multiple roles simultaneously including being sources, destinations and/or  relays.}
\rvn{This  framework is centralized, in the sense that  the design is performed by  a central entity that is aware of the network parameters.
Hence, this framework can be seen as a benchmark for distributed and potentially less comprehensive designs.}
\rvn{In this framework} each  subcarrier can be reused by multiple links, and the nodes  acting as relays operate in the half-duplex mode, i.e., a node cannot send and receive at the same time on the same subcarrier.
\rvn{The objective of the design is to maximize a weighted-sum of the rates injected and reliably communicated over  the network.}  Weights are assumed to be known \emph{a priori}, but can be adapted over time to account for fairness issues and to maintain a desired quality of service. 



Two instances of networks are considered. In the first instance, \rvn{each subcarrier can be time-shared by multiple links, thereby resulting} in continuous subcarrier scheduling variables. The role of these variables is to determine the fraction of time during which a subcarrier is used over a particular link.
In contrast with the first instance, in the second one \rvn{time-sharing is not allowed and a subcarrier, once assigned to a set of links, is used by those  links throughout the signalling interval.}
This instance results in binary subcarrier schedules, \rvn{which using a particular change of variables, are incorporated in the power allocation constraints. This results in a design problem that is significantly easier than its general counterpart considered in the first instance.}
It is worth noting that the \rvn{first} instance is a generalization not only of the \rvn{second} instance, but also of instances in which frequency-reuse is not permitted~\cite{journal_1}. As such, the framework considered in this instance offers significant performance advantages over currently available designs, but at the expense of increasing dimensionality and design complexity. This instance provides an inherent tradeoff between performance and design complexity; the design complexity can be reduced by restricting the number of links that can reuse a particular subcarrier. 

The optimization problems arising from
the joint design in both instances are
nonconvex and hence difficult to solve.
To overcome this difficulty, a logarithmic transformation is used to cast the original problem in a form that, for all but a few constraints, complies with  the geometric programming~(GP)  standard form~\cite{CTPOJ07}. The constraints that are not compatible with that form are approximated by monomial expressions that correspond to their first order Taylor expansion around a given initial point~\cite{BKVH07}.
Using an exponential transformation, the resulting approximation can be cast in a convex form. A refinement of this approximation can be obtained by iterative updating of the initial point. In particular, we use the so-called  iterative monomial approximation technique, wherein the solution of one  convex approximation is used as the initial point in the following iteration. Under relatively mild conditions, this technique is
guaranteed to yield a solution of the Karush-Kuhn-Tucker~(KKT) system \rvn{corresponding} to the original problem~\cite{MW78}.
Numerical results show that the designs developed in both instances yield performance that is considerably superior to that of their counterparts in which frequency-reuse is not allowed.

\rvn{In comparison with currently available designs, the ones presented herein are the first to attempt designing  data routes, subcarrier schedules and power allocation jointly when the subcarriers are both time-shared and frequency-reused. In particular,
the contributions in this paper include: 1)
introducing the concept of simultaneous time-sharing and frequency-reuse of subcarriers; 
2) casting the joint design of data routes, subcarrier schedules and power allocation in a framework that is amenable to GP-based optimization;
3) providing \edt{a simplified approach} that enables mustering a considerable portion of the gains offered by the \edt{full} joint design, but with a significantly lower complexity; and
4) developing an efficient polynomial complexity algorithm for the special case in which the subcarriers can be frequency-reused but not time-shared.
This work builds on the results obtained in~\cite{spawc} and~\cite{RGY13}. However, the exposition herein is more comprehensive and includes additional examples, a simplified \edt{approach} and complexity analysis.}

The paper 
is organized as follows.
Section~\ref{se:lit-rvw} provides an overview of currently available  design techniques. 
Section~\ref{se:pro_stm}  explains the system model  and design objective. 
Section~\ref{se:RPC_w_time} considers the joint design of data routes and power allocations
 when time-sharing of subcarriers is allowed. The complementary instance in which time-sharing is not allowed is addressed in  Section~\ref{se:RPC_w/o_time}. 
The complexity of the proposed algorithms are examined in Section~\ref{se:comp}. Numerical results are provided in Section~\ref{se:sim}, and Section~\ref{se:con} concludes the paper.
For completeness, the GP standard form and the monomial approximation technique are provided in Appendix~\ref{se:GP_mon_def}, and the derivation  of the results pertaining to complexity is provided in Appendix~\ref{se:slf-crdn}.

\section{Related Work} \label{se:lit-rvw}
In this section we provide an overview of the  currently available techniques for routing and resource allocation in wireless networks. \rvn{A plethora} of techniques is available for optimizing each aspect in isolation,  but significantly fewer \rvn{ones} consider their optimization jointly.

Resource allocation in wireless networks constitutes the task of determining the power allocated for each transmission and the fraction of time over which a particular subcarrier is assigned to that transmission. Instances in which resource allocation techniques were developed are provided in~\cite{IH12,KHK05,NS08,JL03,HSAB09,CTPOJ07} for various network scenarios. For instance, power allocation techniques for single-carrier cellular systems and \emph{ad hoc}  multicarrier systems were developed in~\cite{IH12} and~\cite{CTPOJ07}, respectively.  To enable more effective utilization of resources, power allocations were optimized jointly with binary-constrained subcarrier schedules. For instance, 
the designs developed in~\cite{KHK05} and~\cite{NS08} rely on the premise that each subcarrier is exclusively used by one node and the solutions obtained therein are potentially suboptimal.
 When the binary constraint on the subcarrier schedules is relaxed \rvn{allowing the subcarriers}  to be time-shared by multiple nodes, the optimal power allocations \rvn{can be shown to be the water-filling ones~\cite{JL03}}; a related problem was considered in~\cite{HSAB09} for a case in which the nodes experience self-noise.

Further improvement 
can be achieved by joint optimization of \rvn{resource allocations and routing~\cite{journal_1,XJB04,KKKB07,JX06,GW10}}. 
For instance,  a method for obtaining jointly optimal  routes and power allocations was developed \rvn{in~\cite{XJB04} for the case} in which the nodes were restricted to use orthogonal channels for their transmissions.
In a complementary fashion, the case in which the power allocations are fixed was considered in~\cite{KKKB07}. Therein,  a  heuristic was developed  for optimizing  the data routes and subcarrier schedules jointly.

Capitalizing on the potential gains of incorporating power allocation jointly with data routing and subcarrier scheduling, the authors considered  a generic network in which the nodes can assume multiple roles at the same time and each subcarrier could be either used exclusively by one link or time-shared by multiple links~\cite{journal_1}.
Although the designs provided in~\cite{journal_1} offer an effective means for exploiting the resources available for the network, these designs restrict the subcarriers to be used exclusively by only one link at any given time instant. Such a restriction may not incur a significant performance loss in tightly coupled networks~\cite{HL09}, but in networks with clustered structures, this restriction can be quite harmful. For unclustered networks, 
frequency-reuse  may result in a substantial increase in the interference levels. However, if properly exploited, frequency-reuse can yield valuable performance gains. The effect of frequency-reuse was considered in single-channel networks in~\cite{JX06} for  the case in which the data rates are restricted to assume discrete values, and in~\cite{GW10} for the case in which the nodes use superposition coding.

In the current work, we will consider the joint optimization of power allocations, subcarrier schedules and data routes  in the design of generic multicarrier networks with frequency-reuse, and with and without time-sharing. As such, these  designs  generalize currently available ones, and will subsequently offer a significant improvement over their performance.


A summary of this review and a comparison to our work is presented in Table~\ref{tb:lit_rew}.
\begin{table*}
\caption{Related Work} \label{tb:lit_rew}
\begin{center}
\begin{tabular}{lccccl}
\hline
                         Platform  &Routing   &Power Allocation&Scheduling&Frequency-reuse&Reference\\
\hline
Multiuser system                             &$\times$    &$\checkmark$         &$\checkmark$  &$\checkmark$  &   \cite{HL09}  \\
\hline
Uplink cellular               &$\times$    &$\checkmark$         &$\times$      &$\checkmark$  & \cite{CTPOJ07}, \cite{IH12} \\
\hline
OFDMA                    &$\times$    &$\checkmark$         &$\checkmark$  &$\times$      & \cite{KHK05}, \cite{NS08}, \cite{JL03},\cite{HSAB09}   \\
\hline
Generic network                              &$\checkmark$&$\checkmark$         &$\times$      &$\times$      &  \cite{XJB04}\\
\hline
Mesh network                                &$\checkmark$&$\times$         &$\checkmark$  &$\times$      &\cite{KKKB07} \\
\hline
Generic network                              &$\checkmark$&$\checkmark$         &$\times$      &$\checkmark$      &  \cite{JX06,GW10}\\
\hline
Generic network                         &$\checkmark$&$\checkmark$         &$\checkmark$  &$\times$      & \cite{journal_1}\\
\hline
Generic network                                 &$\checkmark$&$\checkmark$         &$\checkmark$  &$\checkmark$   &This work \\
 \hline
\end{tabular}
\end{center}
\end{table*}
\section{Problem Statement} \label{se:pro_stm}
\subsection{System Model} \label{se:sys-mdl}
We consider a multicarrier  wireless network of  $N$ nodes, each with one transmit and one receive antenna, and a fixed power budget, $P_n$, $n\in\mathcal{N}\triangleq\{1,2,\cdots,N\}$. \rvn{The network operates over a  frequency-selective broadband channel of bandwidth  $W_0$, which is partitioned into $K$ frequency-flat narrowband channels,
each  of bandwidth $W=\frac{W_0}{K}$.}
 Node are assumed to be capable of simultaneously transmitting, receiving and  relaying  data. 
This assumption is generic, in the sense that constraining some nodes to  perform a subset of tasks can be readily incorporated in the formulations that will be developed hereinafter. For tractability, the nodes are assumed to 
always have data ready for transmission~\cite{XJB04},
and  for practical considerations, the relaying nodes are assumed to operate in a \rvn{multi-hop, rather than cooperative,} half-duplex mode~\cite{KKKB07}.

\begin{figure}[htb]
\centering
\psfrag{s1}[][][3.5]{$s_4^{(1)}$}
\psfrag{s2}[][][3.5]{$s_3^{(2)}$}
\psfrag{s3}[][][3.5]{$s_4^{(2)}$}
\psfrag{s4}[][][3.5]{$s_3^{(1)}$}
\psfrag{s5}[][][3.5]{$s_2^{(1)}$}
\psfrag{s6}[][][3.5]{$s_1^{(2)}$}
\psfrag{d1}[][][3.5]{$s_1^{(1)}=-s_2^{(1)}-s_3^{(1)}-s_4^{(1)}$}
\psfrag{d2}[][][3.5]{$s_2^{(2)}=-s_1^{(2)}-s_3^{(2)}-s_4^{(2)}$}
\psfrag{a}[][][3]{Node 1}
\psfrag{b}[][][3]{Node 2}
\psfrag{c}[][][3]{Node 3}
\psfrag{d}[][][3]{Node 4}
\psfrag{1}[][][3]{Link 1}
\psfrag{2}[][][3]{Link 2}
\psfrag{3}[][][3]{Link 3}
\psfrag{4}[][][3]{Link 4}
\psfrag{5}[][][3]{Link 5}
\psfrag{6}[][][3]{$\ \  $ Link 6}
\psfrag{7}[][][3]{$\ \ \ \ \  $ Link 7}
\psfrag{8}[][][3]{Link 8}
\psfrag{9}[][][3]{Link 9}
\psfrag{10}[][][3]{$\ $ Link 10}
\psfrag{11}[][][3]{Link 11}
\psfrag{12}[][][3]{Link 12}
\resizebox{0.35\textwidth}{!}{\includegraphics*{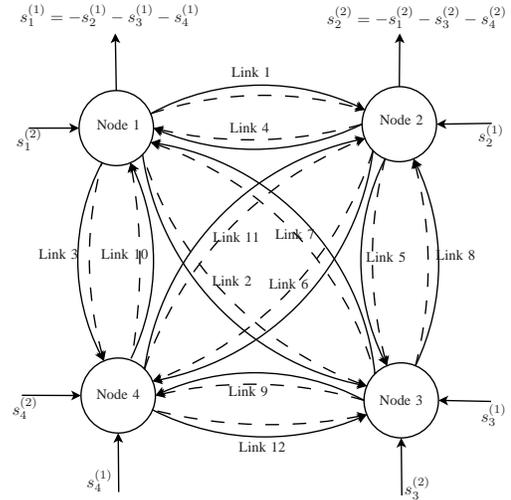}}
\caption{An exemplary network with $N=4$,  $K=2$  and $D=\{1,2\}$. The objective is to maximize a weighted sum of $\{s_n^{(d)}\}$, i.e., the rate injected at node $n$ and intended for destination $d$. }
\label{fig:net}
\end{figure}

The nodes 
are connected with $L$ wireless links, each composed of $K$ subcarriers and 
the set of all links is denoted by $\mathcal{L}\triangleq\{1,2,\cdots,L\}$.
\rvn{The coefficient of the $k$-th subcarrier of link $\ell$ connecting node~$n$ to node~$n^\prime$ is denoted by the complex number $h_{nn^\prime}^{(k)}$ which comprises 
 pathloss, shadowing and fading.}
An instance of such a network with $N=4$ nodes and $K=2$ subcarriers is depicted in Figure~\ref{fig:net}.

In addition to their desired signals, the nodes receive a superposition of noise and interference due to the transmissions of other  nodes in the network. 
Denoting  the  signals transmitted to and received by node $n$ on the $k$-th subcarrier by $u_n^{(k)}$ and  $y_{n}^{(k)}$, respectively,  
we can write \rvn{$y_{n^\prime}^{(k)}=\sum_{n\in\mathcal{N}\setminus\{n^\prime\}}h_{n n^\prime }^{(k)}u_{n}^{(k)}+v_{n^\prime}^{(k)}$,}
where `$\setminus$' denotes the set-minus operation and $v_{n^\prime}^{(k)}$ denotes the corresponding zero-mean additive Gaussian noise with variance $N_0$.
Assuming, as before, that $\ell\in\mathcal{L}$ is the link connecting node~$n$ to node~$n^\prime$,
it can be seen
that  the signal-to-noise-plus-interference ratio~(SNIR) observed by node $n^\prime$ on subcarrier $k$ of  link $\ell$ is given by 
\begin{equation}
\text{SNIR}(\ell,k)=\frac{p_{n k}|h_{nn^\prime}^{(k)}|^2}{WN_0+\sum_{n^{\dprime}\in\mathcal{N}\setminus\{n,n^\prime\}}
p_{n^\dprime}^{(k)}|h_{n^{\dprime}n^\prime}^{(k)}|^2}, \label{eq: SNIR}
\end{equation}
where $p_{n k}$ is the power allocated by node $n$ to the $k$-th subcarrier.
The second term in the denominator of~\eqref{eq: SNIR} represents the aggregate interference observed by node $n^\prime$ on subcarrier~$k$ of  link~$\ell$.
When the nodes transmit Gaussian distributed signals, the maximum data rate that can be reliably communicated on this  subcarrier  is given by $W\log_2(1+\text{SNIR}(\ell,k))$. 

\rvn{For ease of exposition, we divide both the numerator and denominator of~\eqref{eq: SNIR} by $WN_0$ and we use  $g_{\ell k}$ to denote the normalized channel gain,  $\frac{|h_{nn^\prime}^{(k)}|^2}{WN_0}$,  
between any two nodes $n, n^\prime\in\mathcal{N}$.}
\subsection{Network Topology} \label{se:net_top}
The considered network can be represented by a fully-connected weighted directed graph with $N$ vertices and $L=N(N-1)$ links.
To facilitate enumeration of links, the link from node $n$ to node $n^\prime$  will be labelled by $\ell=(N-1)(n-1)+n^\prime-1$ if $n<n^\prime$ and by $ \ell=(N-1)(n-1)+n^\prime$ if $n>n^\prime$.
\rvn{The sets of incoming and outgoing links of node $n\in\mathcal{N}$ are denoted by $\mathcal{L}_-(n)$ and $\mathcal{L}_+(n)$, respectively,} and the connectivity of this graph can be captured by an incidence matrix, $A=[a_{n\ell}]$, \rvn{where $a_{n \ell}=1$ if $\ell\in \mathcal{L}_+(n)$, $a_{n \ell}=-1$ if $\ell \in \mathcal{L}_-(n)$ and $a_{n \ell}=0$ otherwise~\cite{XJB04}.}

\subsection{Design Objective}  \label{se:obj}
Let  $\mathcal{D}\triangleq\{1,\ldots,D\}$
be the set of all destination nodes, where $\mathcal{D}\subseteq \mathcal{N}$.  
Let 
$s_n^{(d)}$ be the  rate of the data stream injected into node $n\in\mathcal{N}$ and intended for destination $d\in\mathcal{D}$.   
The objective of our joint design is to maximize a
weighted-sum of the rates injected into the network, i.e., $\max \sum_{d\in\mathcal{D}}\sum_{n\in\mathcal{N}\setminus\{d\}} w_n^{(d)} s_n^{(d)}$, where
 $\{w_n^{(d)}\}$ are non-negative weights satisfying $\frac{1}{D(N-1)}\sum_{d\in\mathcal{D}}\sum_{n\in\mathcal{N}\setminus \{d\}}w_n^{(d)}=1$.

Assigning  weights to the injected rates provides a convenient means for controlling
the quality of service~(QoS); a higher weight implies a higher priority. 
Weights are typically assigned \emph{a priori}, but can be adapted to meet 
QoS requirements~\cite{HSAB09}. 
Varying the weights enables us to determine the set  of all rates that the proposed design can simultaneously achieve. 

Having described the system model, in Section~\ref{se:RPC_w_time}
we will characterize the constraints that must be satisfied by the routes, the subcarrier schedules, the data rates   and the power allocations. 

\rvn{\section{General Case: Routing and Resource Allocation with Time-sharing} \label{se:RPC_w_time}}
We consider the case when each subcarrier can be both reused and time-shared by multiple links. This case generalizes the cases in which \rvn{either} frequency-reuse or time-sharing of subcarriers \rvn{is} not considered, e.g.,~\cite{journal_1}.
\rvn{After characterizing} the constraints that must be satisfied by the network variables, we  will formulate the cross-layer  design as an optimization problem. Unfortunately this problem is nonconvex and to obtain a \rvn{solution of its}  KKT system, we will use an iterative GP-based technique that 
is guaranteed
to converge to such a \rvn{solution}. 

\subsection{System Constraints}\label{se:RPC_w_time_sys_cons}
In this section, we derive the mathematical constraints that must be satisfied by any feasible set of data routes, time-sharing schedules and power allocations. 
\subsubsection{Routing Constraints} \label{se:RPC_w_time_route_cons}
Let  $x_{\ell k}^{(d)}$ be the data flow  intended for destination $d\in\mathcal{D}$ on subcarrier $k\in\mathcal{K}$ of link $\ell\in\mathcal{L}$. 
The flows, $\{x_{\ell k}^{(d)}\}$,  and the injected rates, $\{s_n^{(d)}\}$, are related by the
flow conservation law, which must be satisfied  at each node.  
This law stipulates that the sum of flows  intended for any destination $d\in\mathcal{D}$ at each node must be equal to zero~\cite{XJB04}.
Applying this law to the current network and using the incidence matrix \rvn{in Section~\ref{se:net_top}}, it can be seen that $\{x_{\ell k}^{(d)}\}$  and $\{s_n^{(d)}\}$ must satisfy the following  constraints: 
\begin{equation}
\sum_{\ell\in\mathcal{L}}\sum_{k\in\mathcal{K}}a_{n l}x_{\ell k}^{(d)}=s_n^{(d)}, \quad\quad\ n\in\mathcal{N}\setminus\{d\},\,d\in\mathcal{D}. \label{eq:flow-cons}
\end{equation}
The flow conservation law implies that the rate of data leaving the network at $d\in\mathcal{D}$  equals the sum of the data rates injected into the network  and intended for this destination. Hence, we can write 
$s_d^{(d)}=-\sum_{n\in\mathcal{N}\setminus\{d\}}s_n^{(d)}$. 
The injected rates,  $\{s_n^{(d)}\}_{n\neq d}$, are non-negative, and since the network is represented by a directed graph, the flows, $\{x_{\ell k}^{(d)}\}$, must be also non-negative. Hence,
\begin{align}
&s_n^{(d)} \geq 0,\qquad\qquad\  n\in\mathcal{N}\setminus\{d\},\,\ d\in\mathcal{D},\label{eq:non-neg-inrate}\\
& x_{\ell k}^{(d)} \geq 0, \qquad\qquad\ \ell\in\mathcal{L},\, k\in\mathcal{K},\, d\in\mathcal{D}. \label{eq:non-neg-flow}
\end{align}

\subsubsection{Scheduling Constraints} \label{se:RPC_w_time_sch_cons}
\begin{figure}[htb]
\centering
\psfrag{1}[][][3.5]{$\ell_1^{(1)}$}
\psfrag{2}[][][3.5]{$\ell_2^{(1)}$}
\psfrag{3}[][][3.5]{$\ell_3^{(1)}$}
\psfrag{k}[][][3.5]{$\begin{matrix}  \text{Signalling interval for} $k=1$\\ \\ \end{matrix}$}
\psfrag{a}[][][3.5]{$\ $ active}
\psfrag{b}[][][3.5]{$\ $ silent}
\psfrag{c}[][][3.5]{$\begin{matrix}\gamma_1^{(1)}(50\%)\\ \end{matrix}$}
\psfrag{d}[][][3.5]{$\begin{matrix}\gamma_{1,2}^{(1)}(20\%)\\ \end{matrix}$}
\psfrag{e}[][][3.5]{$\begin{matrix}\gamma_{1,2,3}^{(1)}(30\%)\\ \end{matrix}$}
\resizebox{0.4\textwidth}{!}{\includegraphics*{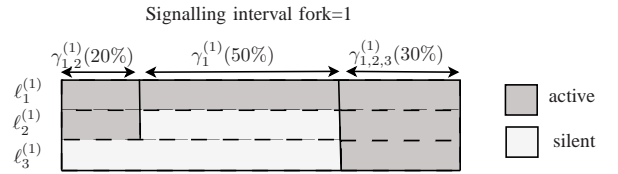}}
\caption{An exemplary scheduling table  for a network with $L=3$,  $K=1$. }
\label{fig:sch}
\end{figure}

Considering both time-sharing and frequency-reuse requires \rvn{introducing a set of variables to characterize the fraction of time over which a particular subset of links utilize the same subcarrier.}
To do so,
\rvn{let} $\gamma_{\ell_1  \cdots \ell_m}^{(k)}$ \rvn{be} the fraction of the signalling interval during which links~$\ell_1,\ldots,\ell_m\in\mathcal{L}$ are simultaneously  `active' on subcarrier $k\in\mathcal{K}$; the remaining $L-m$  links in $\mathcal{L}$ are `silent' on this subcarrier.
Without loss of generality, we will write the indices in an ascending order, i.e., $\ell_1 < \cdots < \ell_m$. For notational convenience, let  $\Gamma$ be the set of all the subcarrier time-sharing schedules. \edt{The cardinality of  $\Gamma$ is given by
\(|\Gamma|=K\sum_{i=1}^L {L \choose i}=K(2^L-1).\)}
\rvn{For instance,}  consider a network with $L=3$ links and $K=1$ subcarriers. \rvn{In this case,} $\Gamma=\{\gamma_1^{(1)},\gamma_2^{(1)},\gamma_3^{(1)},\gamma_{1,2}^{(1)},\gamma_{1,3}^{(1)},\gamma_{2,3}^{(1)},\gamma_{1,2,3}^{(1)}\}$  \rvn{and $|\Gamma|=7$. To see the role of $\Gamma$, }
consider the schedules in Figure~\ref{fig:sch}.  In this figure, $\gamma_1^{(1)}=0.5$, $\gamma_{1,2}^{(1)}=0.2$, $\gamma_{1,2,3}^{(1)}=0.3$, and all the other elements in $\Gamma$ are zero.

Note that the fact that the channels are assumed constant over the signalling interval implies that only the time-sharing schedules (i.e., entries of $\Gamma$) affect the rate expressions, irrespective of the particular time interval over which the subcarriers are time-shared. In other words, horizontal displacement of the shaded blocks in Figure~\ref{fig:sch} does not affect the rate expressions.

\edt{
The number of variables in $\Gamma$ grows exponentially with the number of links, $L$. This renders the incorporation of $\Gamma$ in the joint optimization computationally prohibitive.
In most cases this complexity can be significantly reduced without incurring heavy performance losses. For instance, if the network is tightly coupled, high interference levels render the reuse of subcarriers on multiple links less beneficial. In such a case, restricting the reuse of a subcarrier to  a fewer links may incur negligible deterioration in performance but reduces the number of variables significantly.
To take advantage of this observation, we limit the number of links
that can reuse a particular subcarrier to \edt{$I\ll L$}. By performing this restriction, the number of elements in $\Gamma$ is reduced from $K\bigl(2^L-1\bigr)$ to $K\sum_{i=1}^{I}{L\choose i}$, which, for small  $I$, is polynomial in $L$. For instance, if at most two links are allowed to reuse a particular  subcarrier at any given time, i.e., $I=2$, the number of elements in $\Gamma$ reduces to $\frac{L(L+1)}{2}$.
It is worth noting that
limiting the number of simultaneous transmissions, $I$, inherently offers a trade-off between the performance and complexity. In particular, as
 $I$ increases, the available resources are utilized more efficiently. 
However, our simulations suggest that most of the gain of time-sharing and frequency-reuse is accrued by only considering $I\leq3$ simultaneous
transmissions.
}

\rvn{For} feasible time-sharing schedules, the elements in $\Gamma$ must be non-negative and, \rvn{to ensure no} overlapping in time, 
the total time over which any subcarrier $k\in\mathcal{K}$ is used  must not exceed the length of the signalling interval. These constraints imply that
\begin{align}
&\Gamma \geq 0, &\text{elementwise}, \label{eq:pos-gamma} \\
&\sum_{m=1}^I  \sum_{\ell_1\cdots\ell_m\in\mathcal{L}}  \gamma_{\ell_1 \ldots \ell_m}^{(k)} \leq 1, &\forall k\in\mathcal{K}. \label{eq:time-sum}
\end{align}
\rvn{Note that  summations in~\eqref{eq:time-sum} characterize the number of links, $m$, that reuse a particular subcarrier $k$. 
For instance, for the case in which at most two links reuse this subcarrier, the left hand side~(LHS) of~\eqref{eq:time-sum} can be expressed as  $\sum_{\ell_1\in\mathcal{L}} \gamma_{\ell_1}^{(k)}+\sum_{\ell_1\in\mathcal{L}}\sum_{\ell_2\in\mathcal{L}} \gamma_{\ell_1 \ell_2}^{(k)}$.}

\rvn{Nodes cannot broadcast data to multiple} destinations at the same time, \rvn{that is,} at any time instant, node $n$  can have at most one active link on subcarrier $k$. Hence,  the time-sharing schedules \rvn{corresponding to multiple outgoing links} of node $n$ must be zero. This can be represented as
\begin{multline}
a_{n\ell_1}^+a_{n\ell_2}^+ \Bigl( \gamma_{\ell_1 \ell_2}^{(k)}+\sum_{m=3}^I  \sum_{\ell_3\cdots\ell_m\in\mathcal{L}}  \gamma_{\ell_1 \ldots \ell_m}^{(k)}\Bigr) = 0, \\ 
 \ell_1\in\mathcal{L},\ell_2\in\mathcal{L}\setminus\{\ell_1\},k\in\mathcal{K}, \label{eq:brdcst}
\end{multline}
where $a_{n \ell}^+=\max \{0,a_{n\ell}\}$, that is,  $a_{n \ell}^+=1$ if  $\ell\in\mathcal{L}_+(n)$ and zero, otherwise.

To enforce the half-duplex constraint, we must ensure \rvn{that
no} two links, $\ell_1\in\mathcal{L}_-(n)$ and $\ell_2\in\mathcal{L}_+(n)$,  \rvn{can} be active on the same subcarrier $k\in\mathcal{K}$ at the same time.
This implies that
all the time-sharing schedules that correspond to $\ell_1$ and $\ell_2$, i.e., $\gamma_{\ell_1 \ldots \ell_m}^{(k)}, \ m=2,\ldots,I$, must be zero. Since all the entries in $\Gamma$ are non-negative, these constraints can be written as
\begin{multline}
a_{n\ell_1}^+a_{n\ell_2}^- \Bigl( \gamma_{\ell_1 \ell_2}^{(k)} + \sum_{m=3}^I  \sum_{\ell_3\cdots\ell_m\in\mathcal{L}}  \gamma_{\ell_1 \ldots \ell_m}^{(k)} \Bigr) = 0, \\ 
 \ell_1\in\mathcal{L},\ell_2\in\mathcal{L}\setminus\{\ell_1\},k\in\mathcal{K}, \label{eq:hlf-dplx}
\end{multline}
where $a_{n \ell}^-=|\min \{0,a_{n\ell}\}|$, that is, $a_{n \ell}^-=-1$ if $\ell\in\mathcal{L}_-(n)$ and zero, otherwise. 
Note that~\eqref{eq:brdcst} and~\eqref{eq:hlf-dplx} take effect only when $a_{n\ell}^+a_{n\ell^\prime}^-\neq 0$ and $a_{n\ell}^+a_{n\ell^\prime}^+\neq 0$, respectively.

\subsubsection{Power Allocation Constraints} \label{se:RPC_w_time_power_cons}
\rvn{To facilitate the design, we
replace the node power variables $\{p_{nk}\}$ with
 link power variables $\{q_{\ell k}\}$}, \rvn{which} are related by the following transformation:
\begin{equation}
p_{n k}=\max_{\ell\in\mathcal{L}_+(n)} q_{\ell k}, \qquad\qquad\qquad
 n\in\mathcal{N},\, k\in\mathcal{K}. \label{eq:q_max}
\end{equation}

To gain a better understanding of the transformation in~\eqref{eq:q_max}, we note \rvn{that~\eqref{eq:brdcst}} implies that, of all the links in $\mathcal{L}_+(n)$, only one element  in the set $\{q_{\ell k}\}_{\ell\in\mathcal{L}_+(n)}$, $\forall\; n\in\mathcal{N},\, k\in\mathcal{K},$   can assume a strictly positive value. Now,~\eqref{eq:q_max} indicates that this value is the power allocated by node $n$ to subcarrier  $k$. 
\rvn{Using~\eqref{eq:q_max}, we will formulate our design} in terms of $\{q_{\ell k}\}$ instead of  $\{p_{n k}\}$. These variables must satisfy the following non-negativity constraints:
\begin{equation}
q_{\ell k}\geq0, \quad\quad\quad\quad\  \ell\in\mathcal{L},\,\,k\in\mathcal{K}. \label{eq:non-neg-pow}
\end{equation}

In a practical network, the nodes are likely to have \rvn{individual}  power budgets which bounds the total power used by each node on all  subcarriers. \rvn{To capture} this constraint, we note that only the subcarriers scheduled to outgoing links contribute to the power consumption of each node. More specifically, if $\ell_1\in\mathcal{L}_+(n)$, then all the time-sharing schedules that correspond to $\ell_1$ contribute to the power consumption at node $n$. This constraint can be written as
\begin{multline}
\sum_{k\in\mathcal{K}}\sum_{\ell_1\in\mathcal{L}_+(n)} q_{\ell_1 k}\Bigl( \gamma_{\ell_1}^{(k)}+\sum_{m=2}^I \sum_{\ell_2\cdots\ell_m\in\mathcal{L}} \gamma_{\ell_1 \ldots \ell_m}^{(k)}\Bigr)\leq P_n, \\  n\in\mathcal{N}. \label{eq:pow-bdg_time}
\end{multline}


\subsubsection{Capacity Constraints} \label{se:RPC_w_time_cap_cons}
\new{
To complete the characterization of the network, we point out that  the  data flows and the power allocations are coupled by the maximum aggregate rate that can be supported by the subcarriers of each link. In particular,
the aggregate rate
 $\sum_{d\in\mathcal{D}} x_{\ell k}^{(d)}$  must  not exceed the capacity of  the $k$-th subcarrier of link $\ell$. 

To characterize the capacity constraints, we note that the transmission on  link $\ell\in\mathcal{L}$ and subcarrier $k\in\mathcal{K}$ is composed of two parts. The first part accounts for the fraction of time over which this transmission is interference-free, whereas  
the second part accounts for the fraction of time over which this transmission interferes with other transmissions. To characterize the second part, we identify the interfering links and the fraction of time over which  these links are interfering. To do so, we note that, if subcarrier $k$ is time-shared by links $\ell_1,\ldots,\ell_m$, then the transmissions on links $\ell_2,\ldots,\ell_m$ interfere with the transmission on link $\ell_1$. Hence, the SNIR expression for the transmission on link $\ell_1$ is $\frac{q_{\ell_1 k}g_{\ell_1 k}}{1+\sum_{i=2}^m q_{\ell_i k}g_{\ell^\prime_i k}}$, where $\ell_i^\prime$  denotes the index of the link connecting the  node at which  link $\ell_i$ originates to the  node at which link $\ell_1$ ends.
Since links $\ell_1,\ldots,\ell_m$ are simultaneously active on subcarrier $k$ for a fraction of $\gamma_{\ell_1\ldots\ell_m}^{(k)}$, the expression for the data rate that can be communicated over link $\ell_1$ is
$\gamma_{\ell_1\ldots\ell_m}^{(k)} \log_2 \left(1+\frac{q_{\ell_1 k}g_{\ell_1 k}}{1+\sum_{i=2}^m q_{\ell_i k}g_{\ell^\prime_i k}}\right)$.
Summing over  all possible combinations of the interfering links, the capacity constraint on the aggregate flow of link $\ell_1$ on subcarrier $k$ can be expressed as }
\begin{multline}
\sum_{d\in\mathcal{D}} x_{\ell_1 k}^{(d)} \leq \gamma_{\ell_1}^{(k)} \log_2 (1+q_{\ell_1 k} g_{\ell_1 k})+ \\
\!\!\!\sum_{m=2}^I \sum_{\ell_2\ldots\ell_m\in\mathcal{L}} \!\!\gamma_{\ell_1\ldots\ell_m}^{(k)} \log_2 \left(\!1+\frac{q_{\ell_1 k}g_{\ell_1 k}}{1+\sum_{i=2}^m q_{\ell_i k}g_{\ell^\prime_i k}}\right). 
\label{eq:cap_time}
\end{multline}

\subsection{Problem Formulation} \label{se:RPC_w_time_prob}
To ensure the feasibility of the rates generated by our design, the 
constraints in~\eqref{eq:flow-cons}--\eqref{eq:cap_time} must be satisfied.
Combining these constraints yields the following optimization problem:
\begin{align}
&\max_{\{s_n^{(d)}\},\{x_{\ell k}^{(d)}\},\{q_{\ell k}\},\Gamma}\quad   \sum_{d\in\mathcal{D}}\sum_{n\in\mathcal{N}\setminus\{d\}} w_n^{(d)}s_n^{(d)}, \nonumber\\
&\nonumber \text{subject to}\\
&\quad\text{Routing constraints in~\eqref{eq:flow-cons}--\eqref{eq:non-neg-flow}},\nonumber\\
&\quad\text{Scheduling constraints in~\eqref{eq:pos-gamma}--\eqref{eq:hlf-dplx}},\nonumber\\
&\quad\text{Power allocation constraints in~\eqref{eq:non-neg-pow} and~\eqref{eq:pow-bdg_time}},\nonumber\\
&\quad\text{Capacity constraints in~\eqref{eq:cap_time}}.\label{eq:crs-lyr}
\end{align}
The optimization problem in~\eqref{eq:crs-lyr} is nonconvex because of the
power allocation constraint in~\eqref{eq:pow-bdg_time} and
the capacity constraints in~\eqref{eq:cap_time}.
Examining~\eqref{eq:crs-lyr} reveals that this problem shares some features with the GP standard form, cf.  Appendix~\ref{se:GP_def}.
To exploit this observation,
in the next section we will  perform a  change of variables that will enable us to 
express the objective and all, but one set, of the constraints in a GP-compatible form. The residual  constraints that do not comply with the GP standard form are approximated using the monomial approximation technique in Appendix~\ref{se:mon_def}.
Under relatively mild conditions~\cite{MW78}, iterative application of this technique 
is known to yield  a solution of the KKT system corresponding to~\eqref{eq:crs-lyr}, see e.g.,~\cite{GW10,journal_1}. 

\subsection{\edt{Generalized GP-Based Algorithm}} \label{se:RPC_w_time_GP_alg}
To cast~\eqref{eq:crs-lyr} in a form that is amenable to monomial approximation, we  define two sets of variables, $\{t_n^{(d)}\}$ and $\{r_{\ell k}^{(d)}\}$, which are related to $\{s_n^{(d)}\}$ and $\{x_{\ell k}^{(d)}\}$ by the following maps:
\begin{multline}
s_n^{(d)}=\log_2t_n^{(d)},\quad x_{\ell k}^{(d)}=W\log_2r_{\ell k}^{(d)},\\
  n\in\mathcal{N}\setminus\{d\},\, d\in\mathcal{D},\ell\in\mathcal{L},\, k\in\mathcal{K}. \label{eq:log-flow}
 \end{multline}
These maps are bijective, which renders  recovering $\{s_n^{(d)},x_{\ell k}^{(d)}\}$  
straightforward.
The objective and the routing constraints in~\eqref{eq:crs-lyr}
can be cast in a GP-compatible form. In particular, the objective can be expressed as $\prod_{d\in\mathcal{D}}\prod_{n\in\mathcal{N}\setminus\{d\}}\left(t_n^{(d)}\right)^{w_n^{(d)}}$ and the routing constraints can be expressed as
\begin{alignat}{2}
&\prod_{\ell\in\mathcal{L}}\prod_{k\in\mathcal{K}}\bigl(r_{\ell k}^{(d)}\bigr)^{Wa_{n\ell}} = t_n^{(d)},  \qquad& n\in\mathcal{N}\setminus\{d\},\,d\in\mathcal{D}, \label{eq:cons-flow-gp}\\
&  r_{\ell k}^{(d)} \geq 1,  &\ell\in\mathcal{L},\, k\in\mathcal{K},\, d\in\mathcal{D}, \label{eq:non-neg-flow-gp}\\ 
&  t_n^{(d)} \geq 1,   &n\in\mathcal{N}\setminus\{d\},\, d\in\mathcal{D}.\label{eq:non-neg-rate-gp}
\end{alignat}

The non-negativity constraints in~\eqref{eq:pos-gamma} and~\eqref{eq:non-neg-pow} are inherently satisfied in the GP framework.
The constraints in~\eqref{eq:time-sum} and~\eqref{eq:pow-bdg_time} are already in a GP-compatible form. 
We now consider the constraints in~\eqref{eq:brdcst} and~\eqref{eq:hlf-dplx}.
The right hand side~(RHS) of these constraints are zero, which makes them incompatible with the GP framework in Appendix~\ref{se:GP_def}. This problem can be alleviated by constraining their LHS to be less than an arbitrary small number $\epsilon>0$, \rvn{i.e.,}
\begin{align}
 \nonumber& a_{n\ell_1}^+a_{n\ell_2}^+ \Bigl( \gamma_{\ell_1 \ell_2}^{(k)} +\sum_{m=3}^I  \sum_{\ell_3\cdots\ell_m\in\mathcal{L}}  \gamma_{\ell_1 \ldots \ell_m}^{(k)}\Bigr)\leq \epsilon, \\
& \qquad\qquad\qquad\qquad\qquad \ell_1\in\mathcal{L},\ell_2\in\mathcal{L}\setminus\{\ell_1\},k\in\mathcal{K}, \label{eq:brdcst-epsilon}\\ 
\nonumber& a_{n\ell_1}^+a_{n\ell_2}^- \Bigl( \gamma_{\ell_1 \ell_2}^{(k)} +\sum_{m=3}^I  \sum_{\ell_3\cdots\ell_m\in\mathcal{L}}  \gamma_{\ell_1 \ldots \ell_m}^{(k)}\Bigr) \leq \epsilon,\\
& \qquad\qquad\qquad\qquad\qquad \ell_1\in\mathcal{L},\ell_2\in\mathcal{L}\setminus\{\ell_1\},k\in\mathcal{K}. \label{eq:hlf-dplx-epsilon}
\end{align}

The remaining constraints that are not GP-compatible are those in~\eqref{eq:cap_time}.
 Invoking the change of variables in~\eqref{eq:log-flow}, for $\ell_1 \in \mathcal{L}$ and $k\in\mathcal{K}$, those constraints can be expressed as
 \begin{multline}
\prod_{d\in\mathcal{D}} r_{\ell_1 k}^{(d)} \leq (1+q_{\ell_1 k}g_{\ell_1 k})^{\gamma_{\ell_1}^{(k)}} \times\\
 \prod_{m=2}^I  \prod_{\ell_2\cdots\ell_m\in\mathcal{L}}  \left(1+\frac{q_{\ell_1 k}g_{\ell_1 k}}{1+\sum_{i=2}^m q_{\ell_i k}g_{\ell^\prime_i k}}\right)^{\gamma_{\ell_1\ldots\ell_m}^{(k)}}. \label{eq:log_cap}
 \end{multline}
The RHS of~\eqref{eq:log_cap} is   amenable to
the monomial approximation  technique described in Appendix~\ref{se:mon_def}~\cite{BKVH07}.
One approach to use this technique is to approximate all the terms in the RHS of~\eqref{eq:log_cap} by one monomial. This approach is overly complicated, and an alternative is to approximate each term  by a monomial. The product of these monomials constitutes
a monomial approximation of the RHS of~\eqref{eq:log_cap}. Hence, the constraint in~\eqref{eq:log_cap} can be approximated with 
\begin{multline}
\prod_{d\in\mathcal{D}} r_{\ell_1 k}^{(d)} \leq \mathcal{M}\Bigl((1+q_{\ell_1 k}g_{\ell_1 k})^{\gamma_{\ell_1}^{(k)}}\Bigr)  \times\\
\prod_{m=2}^I \prod_{\ell_2\cdots\ell_m\in\mathcal{L}}
 \mathcal{M}\biggl(  \Bigl(1+\frac{q_{\ell_1 k}g_{\ell_1 k}}{1+\sum_{i=2}^m q_{\ell_i k}g_{\ell^\prime_i k}}\Bigr)^{\gamma_{\ell_1\ldots\ell_m}^{(k)}}\biggr),
\label{eq:cap_monAprx}
 \end{multline}
where the functional $\mathcal{M}(\cdot)$ is described in Appendix~\ref{se:mon_def}. Note that, $\{\gamma_{\ell_1\cdots\ell_m}^{(k)}\}$ are variables and hence inseparable from the argument of $\mathcal{M}(\cdot)$. 

Now, the
problem in~\eqref{eq:crs-lyr} can be approximated by the following GP:
\begin{align}
&\max_{\{t_n^{(d)}\},\{r_{\ell k}^{(d)}\},\{q_{\ell k}\},\Gamma}\quad \prod_{d\in\mathcal{D}}\prod_{n\in\mathcal{N}\setminus\{d\}}\left(t_n^{(d)}\right)^{w_n^{(d)}}, \nonumber\\
&\nonumber \text{subject to}\\
&\nonumber\quad\text{Routing constraints in~\eqref{eq:cons-flow-gp}--\eqref{eq:non-neg-rate-gp}},\\
\nonumber &\quad \text{Scheduling constraints in~\eqref{eq:time-sum},~\eqref{eq:brdcst-epsilon} and~\eqref{eq:hlf-dplx-epsilon}},\\ 
\nonumber &\quad \text{Power allocation constraints in~\eqref{eq:pow-bdg_time}},\\
 &\quad \text{Approximate capacity constraints in~\eqref{eq:cap_monAprx}}.\label{eq:GP}
\end{align}

Note that the relaxations in~\eqref{eq:brdcst-epsilon} and~\eqref{eq:hlf-dplx-epsilon} may result in infeasible  subcarrier time-shares that do not satisfy the constraints in~\eqref{eq:brdcst} and~\eqref{eq:hlf-dplx}. To construct  feasible schedules, the elements of $\Gamma$ that are less than or equal to $\epsilon$ are set to zero.
Using a standard exponential transformation, the GP in~\eqref{eq:GP} can be readily transformed into a convex optimization problem which can be solved in polynomial time using 
interior-point methods~(IPMs)~\cite{BKVH07}.
This implies that~\eqref{eq:GP} enables us to efficiently solve~\eqref{eq:crs-lyr} approximately in the neighbourhood of any initial set $\bigl(\{q_{\ell k}^{(0)}\},\Gamma^{(0)}\bigr)$.  

Finding the global solution for the nonconvex problem in~\eqref{eq:crs-lyr} is difficult, whereas solving the approximated problem in~\eqref{eq:GP} is straightforward. To exploit this fact, we incorporate the formulation  in~\eqref{eq:GP} in an iterative algorithm, whereby the output of solving~\eqref{eq:GP} for an initial  point $\bigl(\{q_{\ell k}^{(0)}\},\Gamma^{(0)}\bigr)$ is used as a starting point for the subsequent iteration. 
\rvn{This technique is \rev{usually} referred to  as the single condensation method, e.g.,~\cite{CP07,GW10}, and under relatively mild conditions, its convergence
to a solution of the KKT system corresponding to~\eqref{eq:crs-lyr} is guaranteed~\cite{MW78}.
Since the original design problem is not convex, this system has multiple \edt{local} solutions and the one to which the single condensation method converges depends on the initial point; \edt{some of the local solutions may be global ones.}}
\rvn{A summary of this algorithm is described in Table~\ref{tb:alg}.
\begin{table}
\caption{Successive GP-based Algorithm for Solving~\eqref{eq:GP}} \label{tb:alg}
\begin{center}
\begin{tabular}{l@{}l}
	\hline
1- &$\ $ Let $U^{(0)}=0$.  Set accuracy to  $\delta>0$. \\
2- &$\ $ Choose $I$ and a feasible $\bigl(\{q_{\ell k}^{(0)}\},\Gamma^{(0)}\bigr)$. \\
3- &$\ $ Solve the GP in~\eqref{eq:crs-lyr}. Denote the value of the objective by $U$.\\
4- &$\ $  \textbf{While} $U-U^{(0)}\geq\delta$,\\
&$\ $$\qquad$$\{q_{\ell k}^{(0)}\}\leftarrow\{q_{\ell k}\}$,\\
&$\ $$\qquad$$U^{(0)}\leftarrow U$,\\
&$\ $$\qquad$Solve the GP in~\eqref{eq:crs-lyr}. Denote the value of the objective by $U$,\\
&$\; $  \textbf{End}.\\
5- &$\ $Remove the elements in $\Gamma$ that are less than $\epsilon$.\\
6- &$\ $Use~\eqref{eq:log-flow} to recover $\{s_n^{(d)}\}$ and $\{x_{\ell k}^{(d)}\}$.\\
 \hline
\end{tabular}
\end{center}
\end{table}
}

\edt{In the next section, we will discuss a special case of this algorithm when time-sharing of subcarriers is not allowed. Before we do that, we now provide a brief discussion on the implementation of this algorithm. To begin with, we note that the algorithm in Table~\ref{tb:alg} is centralized, in the sense that the design is performed by a central entity that is aware of the network parameters. The signalling exchange between the nodes and the central entity, required to establish communication in the considered framework, are described as follows. At the beginning of each signalling interval, the central entity prompts the nodes in the network to sequentially broadcast pilot signals of prescribed power levels. Subsequently, each node computes the subcarrier channel gains from all other nodes in the network. There is  total of $LK$ such gains, where $L$ is the number of links and $K$ is the number of subcarriers.  Each node sends these gains along with its destination nodes, if any, and its priority weights to the central entity.
The central entity performs the joint optimization of the power allocations, scheduling parameters and data routes as described in Table~\ref{tb:alg}. It then forwards these decisions to all the nodes, possibly over a dedicated control channel. In particular, the information forwarded by the central entity include 1) the subcarrier index and the time allocated to each transmission. This information is provided by the set $\Gamma$. The cardinality of this set depends on the number of simultaneous transmissions allowed in each subcarrier, $I$. For instance, for $I=2$, $|\Gamma|=LK(L+1)/2$; 2) The power allocated to each transmission. This information is provided by the set  $\{q_{\ell k}\}$ and the cardinality of this set is  $LK$; and 3) The data rates at each transmitting and receiving node in the route of the stream intended for each destination. This information is provided in the set $\{x_{\ell k}^{(d)}\}$ and the cardinality of this set is $LKD$, where $D$ is the number of intended destinations.
}

\rvn{\section{Special Case: Routing and Resource Allocation without Time-sharing} \label{se:RPC_w/o_time}}
In this section, \rvn{we consider  a design problem similar to the one described in Section~\ref{se:RPC_w_time}, but for the case
when time-sharing of subcarriers is not allowed. This corresponds to the a special case in which the entries of $\Gamma$ in~Section~\ref{se:RPC_w_time_sch_cons} are restricted to be binary.
This restriction results in a mixed integer program} which is generally difficult to solve. To overcome this difficulty, we capture the effect of the scheduling variables in the power allocation constraints. We will show that this approach will enable us to develop a design algorithm with  polynomial-complexity.


\subsection{System Constraints}  \label{se:RPC_w/o_time_sys_cons}
\subsubsection{Routing Constraints}  \label{se:route_cons}
\rvn{These constraints are identical to those 
described in~\eqref{eq:flow-cons}--\eqref{eq:non-neg-flow}.}

\subsubsection{Power Allocation Constraints}  \label{se:power_cons}
In characterizing these constraints, we will use the method described in Section~\ref{se:power_cons}  to denote the power allocated for transmission on subcarrier $k$ of link $\ell$ by the variables $\{q_{\ell k}\}$. 
 These variables must satisfy the non-negativity constraints in~\eqref{eq:non-neg-pow} and the power budget constraint. These constraints, using~\eqref{eq:q_max}, can be cast as 
\begin{equation}
\sum_{k\in\mathcal{K}}\sum_{\ell\in\mathcal{L}}a_{n \ell}^+q_{\ell k}\leq P_n,  \quad\quad\qquad n\in\mathcal{N}. \label{eq:pow-bdg}
\end{equation}

\rvn{Similar to the case considered in Section~\ref{se:RPC_w_time},  the nodes cannot simultaneously broadcast to multiple destinations on the same subcarrier. However, this requirement in the current case can be implicitly captured by the allocation of the link powers. In particular,} for any subcarrier $k\in\mathcal{K}$, any node $n\in\mathcal{N}$ and any  two links $\ell_1,\ell_2\in\mathcal{L}_+(n)$, at least $q_{\ell_1 k}=0$ or $q_{\ell_2 k}=0$, i.e., 
\begin{equation}
a_{n \ell_1}^+ a_{n \ell_2}^+q_{\ell_1 k}q_{\ell_2 k}=0, \quad\ \ell_1,\,\ell_2\in\mathcal{L},\,\,k\in\mathcal{K},\,\,n\in\mathcal{N}.\label{eq:non-mulcast_1}
\end{equation}
\rvn{This constraint is significantly less involved than the one in~\eqref{eq:brdcst}.}

\rvn{Similarly, the half-duplex requirement can be captured by
ensuring that, for each node $n\in\mathcal{N}$, if the power on subcarrier  $k$ of $\mathcal{L}_+(n)$  is strictly  positive, then the power allocated  to this subcarrier on all the links in $\mathcal{L}_-(n)$ is zero, and vice versa.  
Hence, the half-duplex requirement  can enforced by the following constraints:}
\begin{equation}
a_{n \ell_1}^-a_{n \ell_2}^+q_{\ell_1 k} q_{\ell_2 k}=0, \quad\ \ell_1,\,\ell_2\in\mathcal{L},\,\,k\in\mathcal{K},\,\,n\in\mathcal{N}.
\label{eq:non-mulcast}
\end{equation}
\rvn{Note that these constraints are simpler than their counterparts in~\eqref{eq:hlf-dplx}. Also,} note that~\eqref{eq:non-mulcast_1} and~\eqref{eq:non-mulcast}  are trivially satisfied if either link $\ell_1$ or  $\ell_2$ are not
 connected to node $n$.

\subsubsection{Capacity Constraints}  \label{se:cap_cons}
\rvn{In this case, the constraints in~\eqref{eq:cap_time} can be readily  seen to reduce to}
\begin{equation}
\sum_d \frac{x_{\ell_1 k}^{(d)}}{W}\leq\log_2 \Bigl(1+\frac{q_{\ell_1 k}g_{\ell_1 k}}{1+\sum_{\ell_2\in\mathcal{L}\setminus\{\ell_1\}}
q_{\ell_2 k}g_{\ell_2^\prime k}}\Bigr). 
\label{eq:Cap}
\end{equation}

\subsection{Problem Formulation} \label{se:RPC_w/o_time_prob}
\rvn{Using the characterization described in Section~\ref{se:RPC_w/o_time_sys_cons}, the design problem can be cast as:}
\begin{align}
&\max_{\{s_n^{(d)}\},\{x_{\ell k}^{(d)}\},\{q_{\ell k}\}}\quad   \sum_{d\in\mathcal{D}}\sum_{n\in\mathcal{N}\setminus\{d\}} w_n^{(d)}s_n^{(d)}, \nonumber\\
&\nonumber \text{subject to}\\
& \quad \nonumber\text{Routing constraints in~\eqref{eq:flow-cons}--\eqref{eq:non-neg-flow}},\\
\nonumber &\quad \text{Power allocation constraints in~\eqref{eq:non-neg-pow},\rvn{~\eqref{eq:pow-bdg}}--\eqref{eq:non-mulcast}},\\
&\quad \text{Capacity constraints in~\eqref{eq:Cap}}.\label{eq: JRPA}
\end{align}

The optimization problem in~\eqref{eq: JRPA} is nonconvex because the RHS of~\eqref{eq:Cap} is the logarithm of a rational function, and therefore not concave. 
The equality constraints in~\eqref{eq:non-mulcast_1} and~\eqref{eq:non-mulcast} are not affine 
and hence,  nonconvex.
 In the next section, we will develop a GP-based algorithm, analogous to the one described in Section~\ref{se:RPC_w_time_GP_alg},
 to obtain a locally optimal solution.

\subsection{Proposed GP-based Algorithm} \label{se:RPC_w/o_time_GP_alg}


The optimization problem in~\eqref{eq: JRPA}, although nonconvex, is amenable to the GP-based monomial approximation in Appendix~\ref{se:mon_def}. To use this approximation,
we use~\eqref{eq:log-flow} to transform $\{s_n^{(d)}\}$ and $\{x_{\ell k}^{(d)}\}$ to  $\{t_n^{(d)}\}$ and $\{r_{\ell k}^{(d)}\}$, respectively.
Using these new variables,
 the routing constraints are readily expressed in a GP-compatible form as described in~\eqref{eq:cons-flow-gp}--\eqref{eq:non-neg-rate-gp}.

Substituting from~\eqref{eq:log-flow} into~\eqref{eq:Cap} yields the following set of equivalent constraints:
\begin{multline}
\Bigl(1 +\!\!\!\sum_{\ell_2\in\mathcal{L}\setminus\{\ell_1\}}q_{\ell_2 k}g_{\ell_2^\prime k}\Bigr)\prod_{d\in\mathcal{D}} r_{\ell_1 k}^{(d)}\leq 1 +\sum_{\ell_2\in\mathcal{L}}q_{\ell_2 k}g_{\ell_2^\prime k},\\
k\in\mathcal{K},\,\ell_1\in\mathcal{L}, \label{eq:non-posy cap}
\end{multline}
which are, unfortunately 
not GP-compatible. Using the monomial approximation technique in Appendix~\ref{se:mon_def} yields the following approximation of~\eqref{eq:non-posy cap}
in the neighbourhood of $\{q_{\ell k}^{(0)}\}$: 
\begin{multline}
\Bigl(1 +\!\!\!\!\!\!\sum_{\ell_2\in\mathcal{L}\setminus\{\ell_1\}}\!\!\!\! q_{\ell_2 k}g_{\ell_2^\prime k}\Bigr)\prod_{d\in\mathcal{D}} r_{\ell_1 k}^{(d)}
\leq c_{\ell_1 k}\!\prod_{\ell_2\in\mathcal{L}}\Bigl({q_{\ell_2 k}}/{q_{\ell_2 k}^{(0)}}\Bigr)^{\theta_{\ell_2 k}},\\
k\in\mathcal{K},\ell_1\in\mathcal{L}, \label{eq:gp_pos}
\end{multline}
 where $\{q_{\ell k}^{(0)}\}$ is the initial power allocation, \(c_{\ell_1 k}=1+ \sum_{\ell_2\in\mathcal{L}}q_{\ell_2 k}^{(0)} g_{\ell_2^\prime k}\),
 and $\theta_{\ell_2 k}=q_{\ell_2 k}^{(0)} g_{\ell_2^\prime k}/c_{\ell_1 k}$.

Analogous to the case considered in Section~\ref{se:RPC_w_time_GP_alg},~\eqref{eq:non-mulcast_1} and~\eqref{eq:non-mulcast} are replaced with the GP-compatible inequality constraints.
The joint design of data routes and power allocations in~\eqref{eq: JRPA} can be approximated with the following GP:
\begin{align}
&\max_{\{t_n^{(d)}\},\{r_{\ell k}^{(d)}\},\{q_{\ell k}\}}\quad \prod_{d\in\mathcal{D}}\prod_{n\in\mathcal{N}\setminus\{d\}}\left(t_n^{(d)}\right)^{w_n^{(d)}}, \nonumber\\
&\nonumber \text{subject to}\\
\nonumber &\quad\!\!\! \text{Routing constraints in~\eqref{eq:cons-flow-gp}--\eqref{eq:non-neg-rate-gp}},\\
\nonumber &\quad\!\!\! \text{Power allocation constraints in~\eqref{eq:pow-bdg}--\eqref{eq:non-mulcast} (relaxed versions)},\\
 &\quad\!\!\! \text{Approximate capacity constraints in~\eqref{eq:gp_pos}}.\label{eq: JRPA-GP}
\end{align}
\rvn{A locally optimal solution of~\eqref{eq: JRPA} can be obtained by solving~\eqref{eq: JRPA-GP} iteratively 
using the single condensation method described in Section~\ref{se:RPC_w_time_GP_alg}.}


\section{Complexity Analysis}  \label{se:comp}
In this section we examine the computational complexity required for solving the problems described in Sections~\ref{se:RPC_w_time_GP_alg} and~\ref{se:RPC_w/o_time_GP_alg} for the cases with and without time-sharing, respectively.
The algorithms in these sections iteratively solve the families of the optimization problems in~\eqref{eq:GP} and~\eqref{eq: JRPA-GP}. \rvn{Being in a GP-compatible form, these problems  can be readily converted into convex forms and can be efficiently solved using IPM-based solvers.}

In IPM, the objective and inequality constraints are used to construct a log-barrier function which is minimized along a central path using Newton's method.
The complexity of each Newton step  grows with the cube of the number of inequality constraints and the number of Newton steps can be bounded if the log-barrier function is self-concordant~\cite{Boyd}, cf. Appendix~\ref{se:slf-crdn}.
In that case, the number of Newton steps can be shown to grow with the square root of the number of inequality constraints~\cite{Boyd}.

Unfortunately, the log-barrier functions related to the problems in~\eqref{eq:GP} and~\eqref{eq: JRPA-GP} are not self-concordant. To circumvent this difficulty,   we introduce  a set of auxiliary variables and constraints which, although redundant,  enables us to construct self-concordant log-barrier functions.  Using these functions and the results in~\cite{Boyd}, we arrive at the following proposition: 
\begin{proposition} \label{pr:GP}
The complexity of solving~\eqref{eq:GP} with IPM-based solvers is of order
\begin{equation*}
\mathcal{O}\left(\left(2LKN+N+D(N-1)+2K\sum_{i=1}^{I}{L\choose i}\right)^{3.5}\right),
\end{equation*}
and the complexity of solving~\eqref{eq: JRPA-GP} with IPM-based solvers is of order
\begin{equation*}
\mathcal{O}\left(\left(LK(3L+2)+N+D(N-1)\right)^{3.5}\right).
\end{equation*}
\end{proposition}
\begin{IEEEproof}
See Appendix~\ref{se:slf-crdn}.
\end{IEEEproof}

The first statement of Proposition~\ref{pr:GP} pertains to the general case with time-sharing and frequency-reuse. This statement shows that the complexity of solving the problem in~\eqref{eq:GP}
\edt{is polynomial in $L$ for small values of $I$.}
\edt{
 The complexity of solving~\eqref{eq:GP} can be further reduced} by combining the broadcasting constraint and the half-duplex constraint in~\eqref{eq:brdcst} and~\eqref{eq:hlf-dplx}, respectively. In particular,  examining these constraints reveals that they are related to the network topology and do not depend on the channel conditions. 
Hence, these two constraints can be enforced by pruning the set $\Gamma$ prior to solving~\eqref{eq:crs-lyr} or its approximated version in~\eqref{eq:GP}. The pruning rule is as follows:  For each $\ell$ and $\ell^\prime\in\mathcal{L}$, if either $a_{n\ell_1}^+a_{n\ell_2}^+=0$ or $a_{n\ell_1}^+a_{n\ell_2}^-=0$, the corresponding time-shares in~\eqref{eq:brdcst} and~\eqref{eq:hlf-dplx} are removed  from the set $\Gamma$. Unfortunately, we have not been able to obtain a closed form of the cardinality of the resulting $\Gamma$. However, the reduction in complexity, at least for small networks, appears to be significant. For instance, for fully
connected networks with $N=4$ nodes and $L=N(N-1)=12$ links, $|\Gamma|$ is reduced from 4095 to 40.

The \edt{second} statement of Proposition~\ref{pr:GP} pertains to the special case in which time-sharing is not allowed. This statement shows that the complexity of solving~\eqref{eq: JRPA-GP}
is polynomial in the number of nodes, $N$, and the number of subcarriers, $K$. In particular, it grows as $L^7K^{3.5}$.
\rvn{Another case in which the design complexity is polynomial is the one in which the subcarriers are time-shared but not frequency-reused~\cite{journal_1}. In that case the design complexity is $\mathcal{O}((LK(4+D)+N+K+D(N-1))^{3.5})$. Hence the special cases with either no frequency-reuse or no time-sharing  have polynomial complexity.}



\section{Simulation Results}  \label{se:sim}
In this section we provide numerical results  to evaluate the performance of joint routing and resource allocation algorithms for the cases with and without time-sharing. 



The locations of the nodes are randomly generated and evenly distributed over a $300\times300\ \text{m}^2$ square.
The nodes are assumed to have identical power budgets, i.e, $P_n=P, \ \forall n\in\mathcal{N}$, and
\rvn{the available frequency-selective channel is partitioned into a set of frequency-flat Rayleigh fading channels}
with the values of pathloss~(PL) and shadowing components obtained from the
non line-of-sight communication of indoor hotspot~(InH) scenario in the IMT-Advanced document~\cite{IMT-Advanced}.
According to~\cite{IMT-Advanced}, the PL component on link $\ell\in\mathcal{L}$ is given by
\begin{equation}
PL=43.3\log_{10}(d_{\ell})+11.5+20\log_{10}(f_c), \label{eq:pathloss}
\end{equation}
where $d_{\ell}$ is the length of link $\ell$ in meters and $f_c$ is the carrier frequency in Gigahertz which, in our simulations, is set to $f_c=3.4$~GHz.
The shadowing component is assumed to be log-normal distributed with a mean of  0 dB and a standard deviation of 4 dB. The Rayleigh fading component is generated by the envelope of a zero-mean unit-variance complex Gaussian-distributed random variable.
The available bandwidth around each subcarrier is set to $W=200$~KHz and the noise power density at receivers is set to $N_0=-174$~dBm/Hz.

The results reported herein are obtained using the  \textsf{CVX} package~\cite{CVX} with an underlying \textsf{MOSEK} solver~\cite{MOSEK}. The value of $\epsilon$ in~\eqref{eq:brdcst} and~\eqref{eq:hlf-dplx}  is
set to $10^{-4}$. 

\begin{figure}        \centering
     \subfigure[]{
                \centering
                \psfrag{a}[][][2.5]{Node 1}
                \psfrag{b}[][][2.5]{Node 2}
                \psfrag{c}[][][2.5]{Node 3}
                \psfrag{d}[][][2.5]{Node 4}
                \psfrag{4}[][][2.8]{0.5 b/s/Hz}
                \psfrag{12}[][][2.8]{2 b/s/Hz}
                \psfrag{7}[][][2.8]{0.5 b/s/Hz}
                \psfrag{11}[][][2.8]{2 b/s/Hz}
                \psfrag{10}[][][2.8]{0.6 b/s/Hz}
                \psfrag{s1}[][][2.8]{$s_4^{(1)}=3.1$ b/s/Hz}
                \psfrag{d1}[][][2.8]{$s_1^{(1)}=3.1$ b/s/Hz}
                \resizebox{.2\textwidth}{!}{\includegraphics[width=\textwidth]{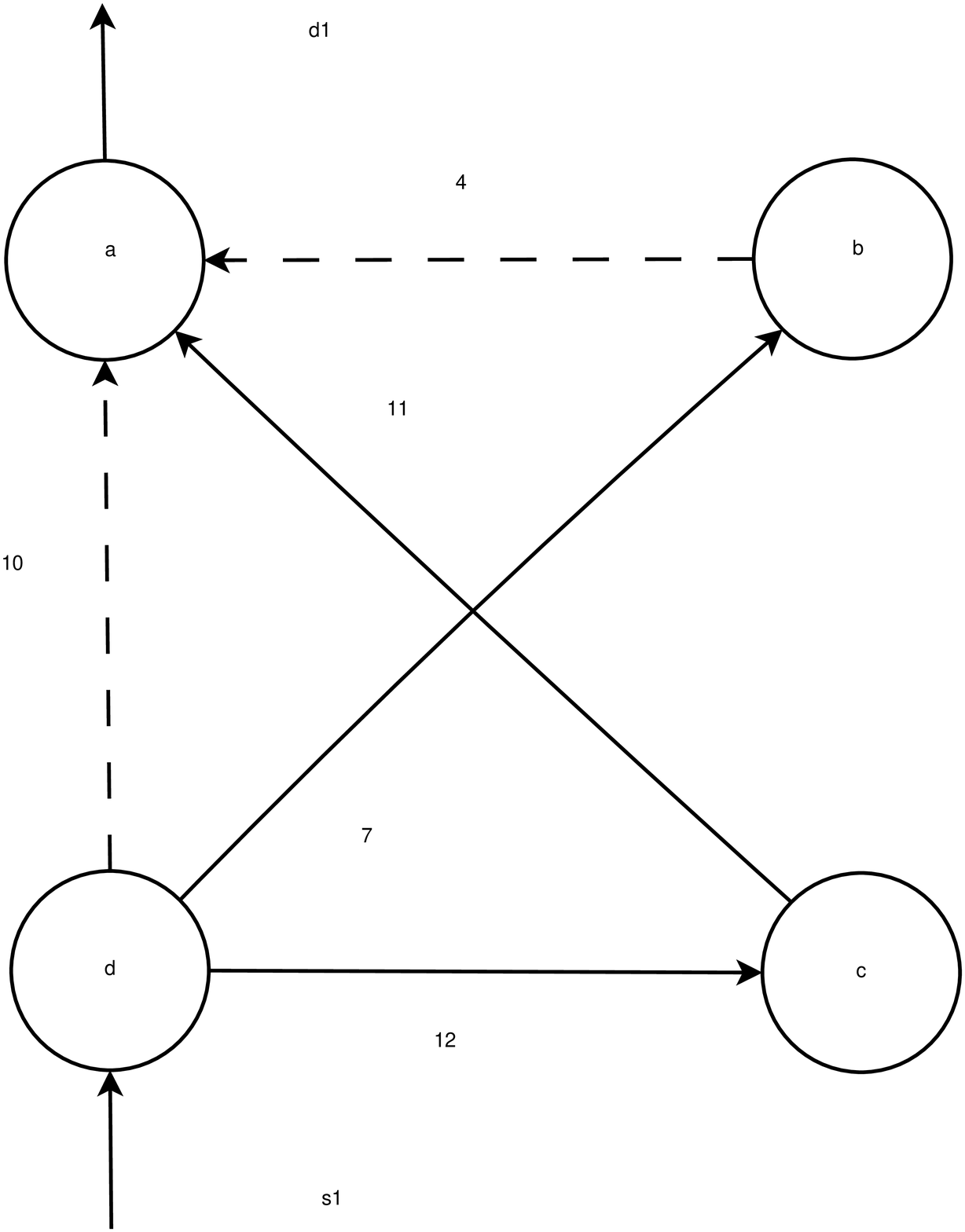}}
              \label{fig:time_1}}
         \subfigure[]{
                \centering
                \psfrag{k1}[][][2.5]{$k=1$}
                \psfrag{k2}[][][2.5]{$k=2$}
                \psfrag{a}[][][2]{Node 1}
                \psfrag{b}[][][2]{Node 2}
                \psfrag{c}[][][2]{Node 3}
                \psfrag{d}[][][2]{Node 4}
                \psfrag{3}[][][2]{2.8 b/s/Hz}
                \psfrag{7}[][][2]{4.3 b/s/Hz}
                \psfrag{11}[][][2]{2.8 b/s/Hz}
                \psfrag{9}[][][2]{1.5 b/s/Hz}
                \psfrag{s1}[][][2]{$s_3^{(2)}=4.3$ b/s/Hz}
                \psfrag{d1}[][][2]{$s_2^{(2)}=4.3$ b/s/Hz}
                \resizebox{.28\textwidth}{!}{\includegraphics[width=\textwidth]{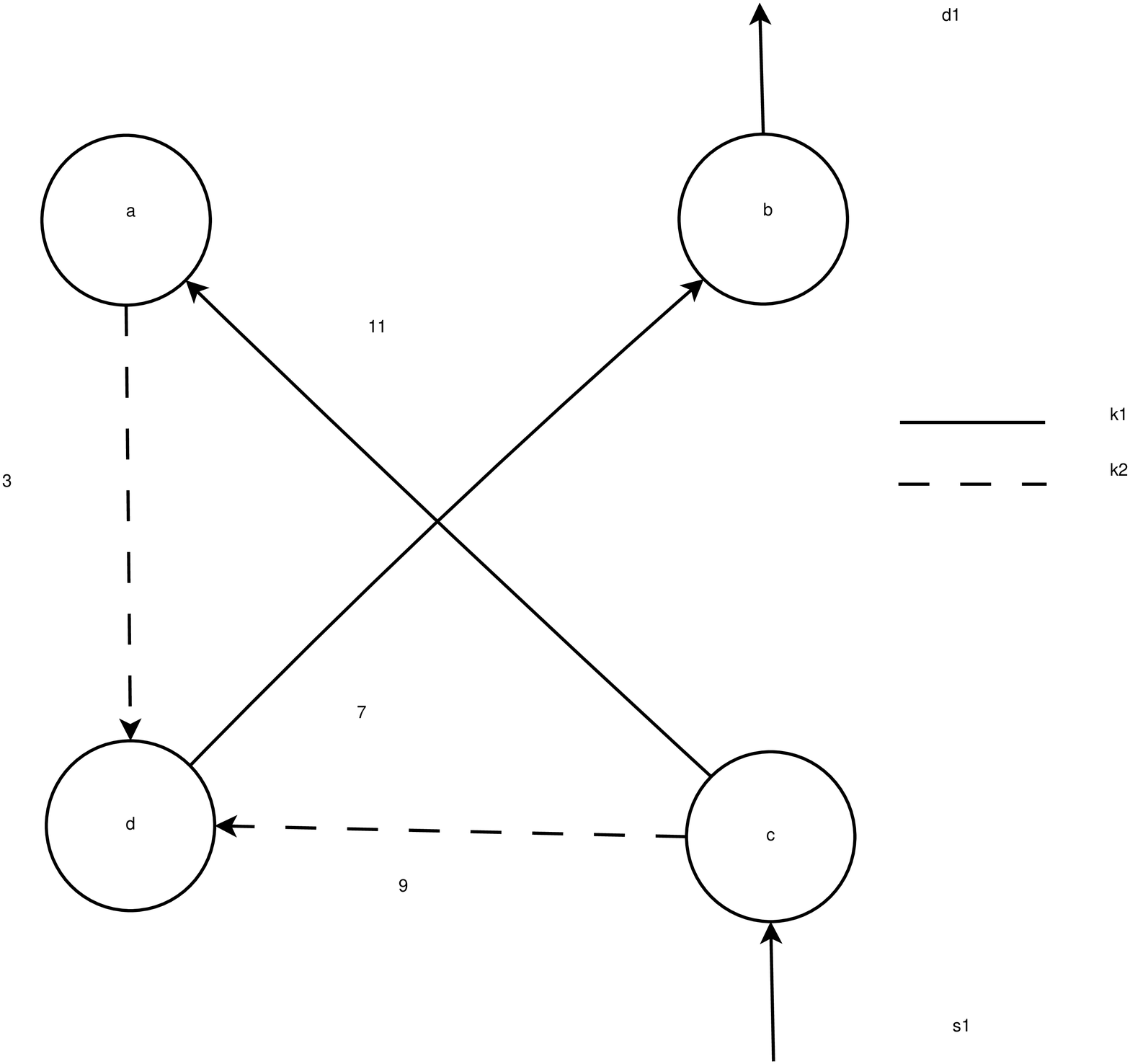}}
                \label{fig:time_2}}
      \vspace{-.2 cm}
      \caption{Data routes for \subref{fig:time_1}  $d=1$, \subref{fig:time_2} $d=2$ in Example~1.}
      \vspace{-.5 cm}
     \label{fig:Time}
\end{figure}

\begin{figure}[ht] \centering
\vspace{-.8 cm}
\psfrag{k1}[][][4]{$k=1$}
\psfrag{k2}[][][4]{$k=2$}
\psfrag{4}[][][3]{Link 4}
\psfrag{12}[][][3]{$\,$ Link 12}
\psfrag{3}[][][3]{Link 3}
\psfrag{7}[][][3]{Link 7}
\psfrag{11}[][][3]{$\,$ Link 11}
\psfrag{100}[][][3]{Link 10}
\psfrag{9}[][][3]{Link 9}
\psfrag{60}[][][2.5]{$60\%$}
\psfrag{10}[][][2.5]{$10\%$}
\psfrag{40}[][][2.5]{$40\%$}
\psfrag{30}[][][2.5]{$30\%$}
\psfrag{45}[][][2.5]{$45\%$}
\psfrag{15}[][][2.5]{$15\%$}
\resizebox{0.48\textwidth}{!}{\includegraphics*{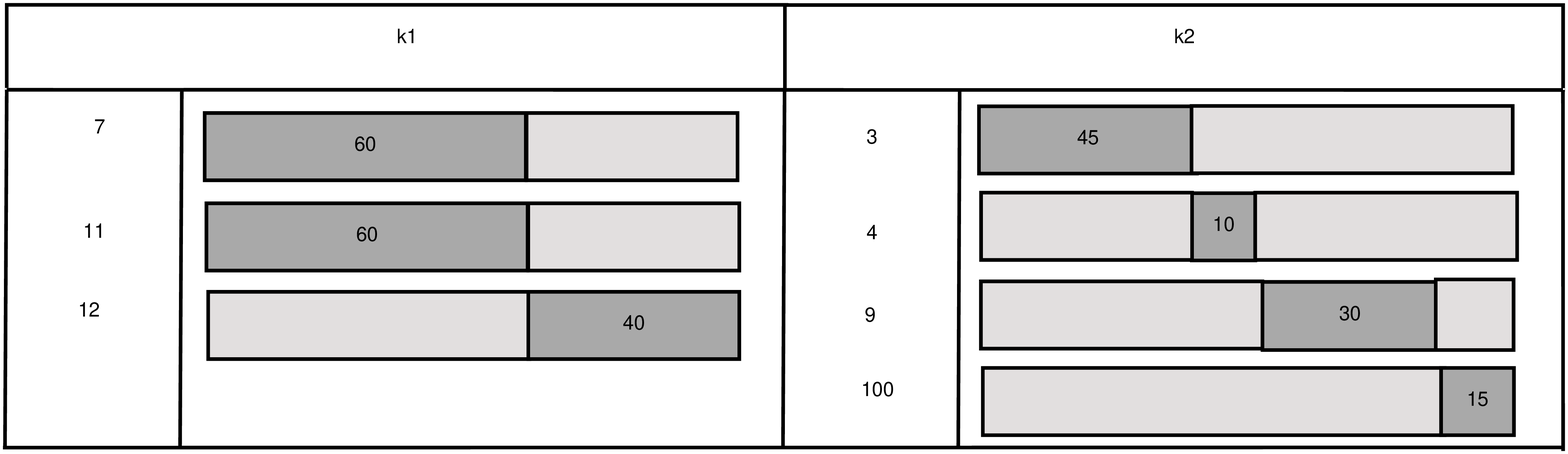}}
\caption{Time-sharing schedules of the subcarriers in Example~1.} \label{tb:Time_share}
\vspace{-.4 cm}
\end{figure}

\textit{Example 1: (Joint Routing and Resource Allocation with Time-sharing)}
Consider an exemplary network with $N=4$ nodes. In this network, nodes 3 and 4 wish to communicate with nodes 2 and 1, respectively, over $K=2$ subcarriers. In particular, for destination node $d=1$, the source is node $n=4$ and nodes $\{2,3\}$ are potential relays,  and, for destination node $d=2$, the source is node $n=3$ and nodes $\{1,4\}$ are potential relays.
The considered network has $L=12$ directional links and therefore the channel matrix has $12\times2$ elements. \rvn{The channels are assumed to be static} and their normalized gain in dB, i.e.,
 $10\log_{10} g_{\ell k}$, is given in Table~\ref{tb:ch_gain}.
\begin{table}
\caption{Normalized Channel Gains, $\{g_{\ell k}\}$, in Example~1 {[dB]}.} \label{tb:ch_gain}
\vspace{-.2 cm}
\begin{center}
\begin{tabular}{c|cccccc}
& link 1& link 2&link 3&link 4&link 5&link 6\\
 \hline
subcarrier 1& -6.1 &  -11.1 &   1.86 & -12.3 & -13.5 &42.4 \\
subcarrier 2& -5.7  & 1.57 &   5.64  & -5.98 & -19.2 &40.7 \\
\hline
& link 7&link 8&link 9&link 10&link 11&link 12\\
\hline
subcarrier 1&      15.82 & -2.43 & -5.82 & -6.9  &  35.0 & 3.304 \\
subcarrier 2&     -3.38 & -5.61 & 0.871 &  -0.6 & 41.2  & -14.5\\
\end{tabular}
\end{center}
\vspace{-0.3 cm}
\end{table}
\new{In this example, the power budget of each node is set to $P=20$~dBm, \edt{the number of simultaneous transmissions is set to $I=3$} and the two rates, $s_3^{(2)}$ and $s_4^{(1)}$, are assigned equal weights, i.e., $w_3^{(2)}=w_4^{(1)}=1$. Since in this example time-sharing is allowed, the algorithm in Section~\ref{se:RPC_w_time_GP_alg} is used to generate the data routes, time-sharing schedules and power allocations.

The sum-rate yielded by the algorithm in Section~\ref{se:RPC_w_time_GP_alg} is 7.4~b/s/Hz. The data routes generated by this algorithm are illustrated in Figure~\ref{fig:Time}. 
For ease of exposition, the network in this example is split into the two sub-networks: the one in Figure~\ref{fig:time_1} depicts the routes of the data intended for destination $d=1$, and the one in Figure~\ref{fig:time_2} depicts the routes of the data intended for destination $d=2$. The complete network is the superposition of the two sub-networks. For instance, the data transmitted over link 7, connecting node 3 to node 1, is 4.8~b/s/Hz, of which 2~b/s/Hz is intended for destination $d=1$ and 2.8 b/s/Hz is intended for destination $d=2$. 

The time-sharing schedules of the subcarriers generated by the algorithm in Section~\ref{se:RPC_w_time_GP_alg} are provided in Figure~\ref{tb:Time_share}. It can be seen from this figure that subcarrier $k=1$ is both reused and time-shared, whereas subcarrier $k=2$ is only time-shared. 
Figures~\ref{fig:Time} and~\ref{tb:Time_share} imply that link 7, connecting node 3 to node 1, and link 11, connecting node 4 to node 2, carry the data intended for both destinations on the same subcarrier, $k=1$, during the same time interval. 
The fact that our designs enforce half-duplex requirement can be inferred from these figures. For instance, Figure~\ref{fig:time_1} shows that node 3 uses the same subcarrier, $k=1$, for its transmission and reception on links 7 and 12, respectively, but Figure~\ref{tb:Time_share} shows that transmission and reception occur during different time intervals. 


The power allocations yielded by the algorithm in Section~\ref{se:RPC_w_time_GP_alg} are shown in Table~\ref{tb:time_power}.
This table shows that, because of frequency-reuse, the nodes do not necessarily use their total power budgets. This is due to the fact that, in this scenario, when a node increases its transmission power, it inflicts high interference on other transmissions. This is in contrast with the situation considered in~\cite{journal_1}, wherein frequency-reuse is not allowed and increasing the transmitted power of a node does not affect the transmissions of the other nodes in the network.}
%
\hfill$\Box$

\begin{table}
\caption{Power Allocations (mW) in Example~1.} \label{tb:time_power}
\vspace{-.2 cm}
\begin{center}
\begin{tabular}{|l|l|l|l|}
\hline
$n=1$ &$n=2$ &$n=3$ &$n=4$ \\
	\hline
$q_{3,2}=23$   &  $q_{4,2}=56.5$ & $q_{7,1}=6.5$  & $q_{10,2}=24.5$ \\
                &                & $q_{9,2}=13.5$  & $q_{11,1}=0.3$ \\
               &                &                 & $q_{12,1}=16.5$ \\
\hline
\end{tabular}
\end{center}
\vspace{-.7 cm}
\end{table}

\textit{Example 2: (Joint Routing and Resource Allocation without Time-sharing)}
Consider an exemplary network with $N=6$ nodes. In this network, as before, nodes 3 and 4 wish to communicate with nodes 2 and 1, respectively, over $K=4$ subcarriers. In particular, for destination node $d=1$, the source is node $n=4$ and the other nodes, i.e., $\{2,3,4,5\}$ are potential relays,  and, for destination node $d=2$, the source is node $n=3$ and nodes $\{1,4,5,6\}$ are potential relays.

The considered network has $L=30$ links and therefore the channel matrix has $30\times4$ elements.
For space considerations, this matrix is not provided, but 
since the channel gain on each subcarrier is dominated by the PL component, we provide the coordinates of the nodes in the $300\!\times\! 300 \ \text{m}^2$ square; calculating the PL components from these coordinates is straightforward, cf.~\eqref{eq:pathloss}. The coordinates of the nodes are $ \{(283,202),(191,208),(287,20),(72,76),(201,67),(86,200)\}$.

\begin{figure}
\centering
     \subfigure[]{
                \centering
                \psfrag{a}[][][1.5]{Node 1}
                \psfrag{b}[][][1.5]{Node 2}
                \psfrag{c}[][][1.5]{Node 3}
                \psfrag{d}[][][1.5]{Node 4}
                \psfrag{e}[][][1.5]{Node 5}
                \psfrag{f}[][][1.5]{Node 6}
                \psfrag{5}[][][2]{4.8 b/s/Hz}
                \psfrag{s}[][][2]{$s_4^{(1)}=4.8$ b/s/Hz}
                \psfrag{y}[][][2]{$s_1^{(1)}=4.8$ b/s/Hz}
                \resizebox{.3\textwidth}{!}{\includegraphics[width=\textwidth]{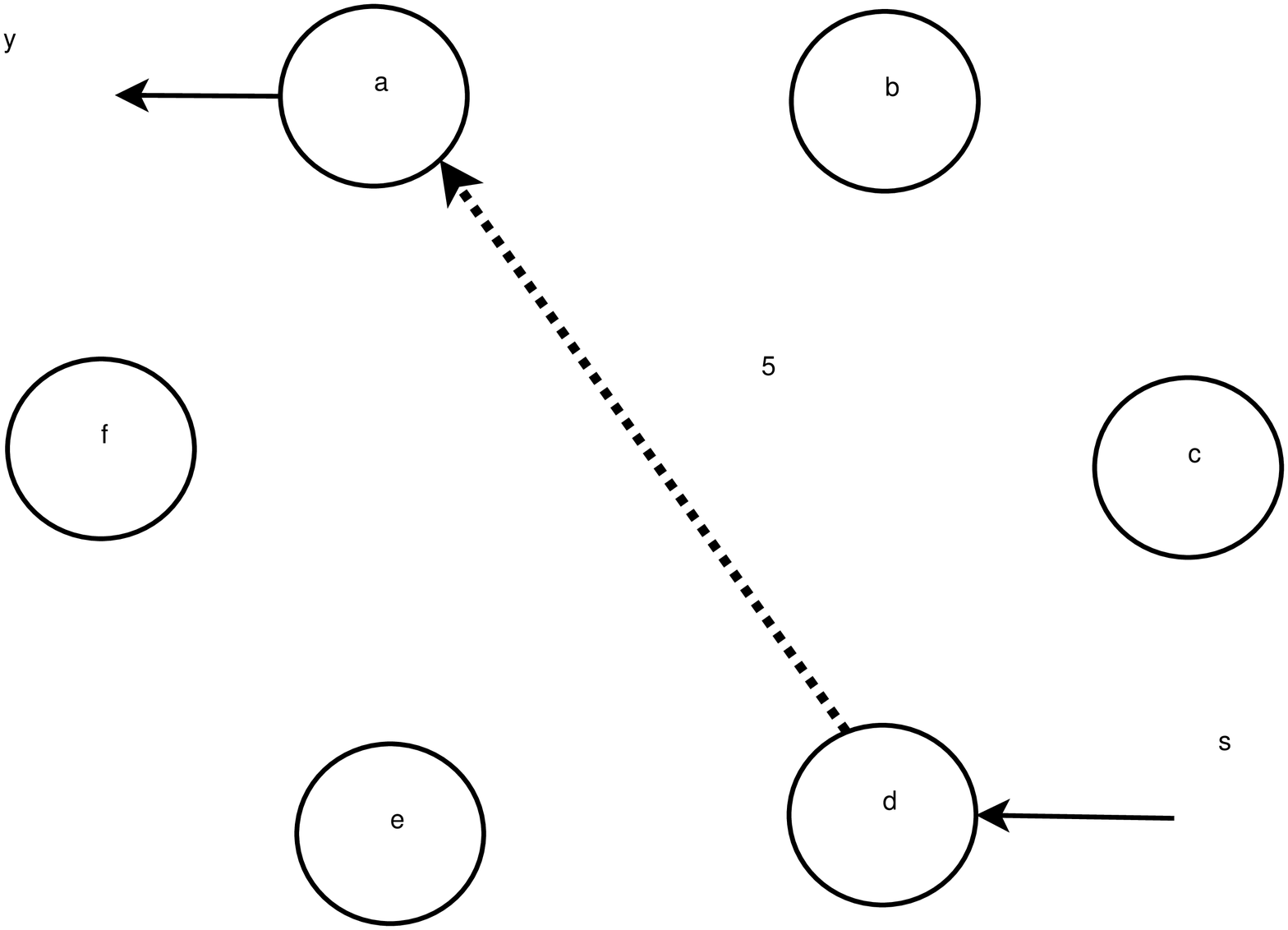}}
              \label{fig:sub_1}}
        \quad
         \subfigure[]{
                \centering
                \psfrag{a}[][][1.1]{Node 1}
                \psfrag{b}[][][1.1]{Node 2}
                \psfrag{c}[][][1.1]{Node 3}
                \psfrag{d}[][][1.1]{Node 4}
                \psfrag{e}[][][1.1]{Node 5}
                \psfrag{f}[][][1.1]{Node 6}
                \psfrag{k1}[][][1.5]{$\ k=1$}
                \psfrag{k2}[][][1.5]{$\ k=2$}
                \psfrag{k3}[][][1.5]{$\ k=3$}
                \psfrag{k4}[][][1.5]{$\ k=4$}
                \psfrag{1}[][][1.5]{5.2 b/s/Hz}
                \psfrag{2}[][][1.5]{5.2 b/s/Hz}
                \psfrag{3}[][][1.5]{3.9 b/s/Hz}
                \psfrag{4}[][][1.5]{3.9 b/s/Hz}
                \psfrag{s}[][][1.5]{$s_3^{(2)}=9.1$ b/s/Hz}
                \psfrag{y}[][][1.5]{$s_2^{(2)}=9.1$ b/s/Hz}
                \resizebox{.4\textwidth}{!}{\includegraphics[width=\textwidth]{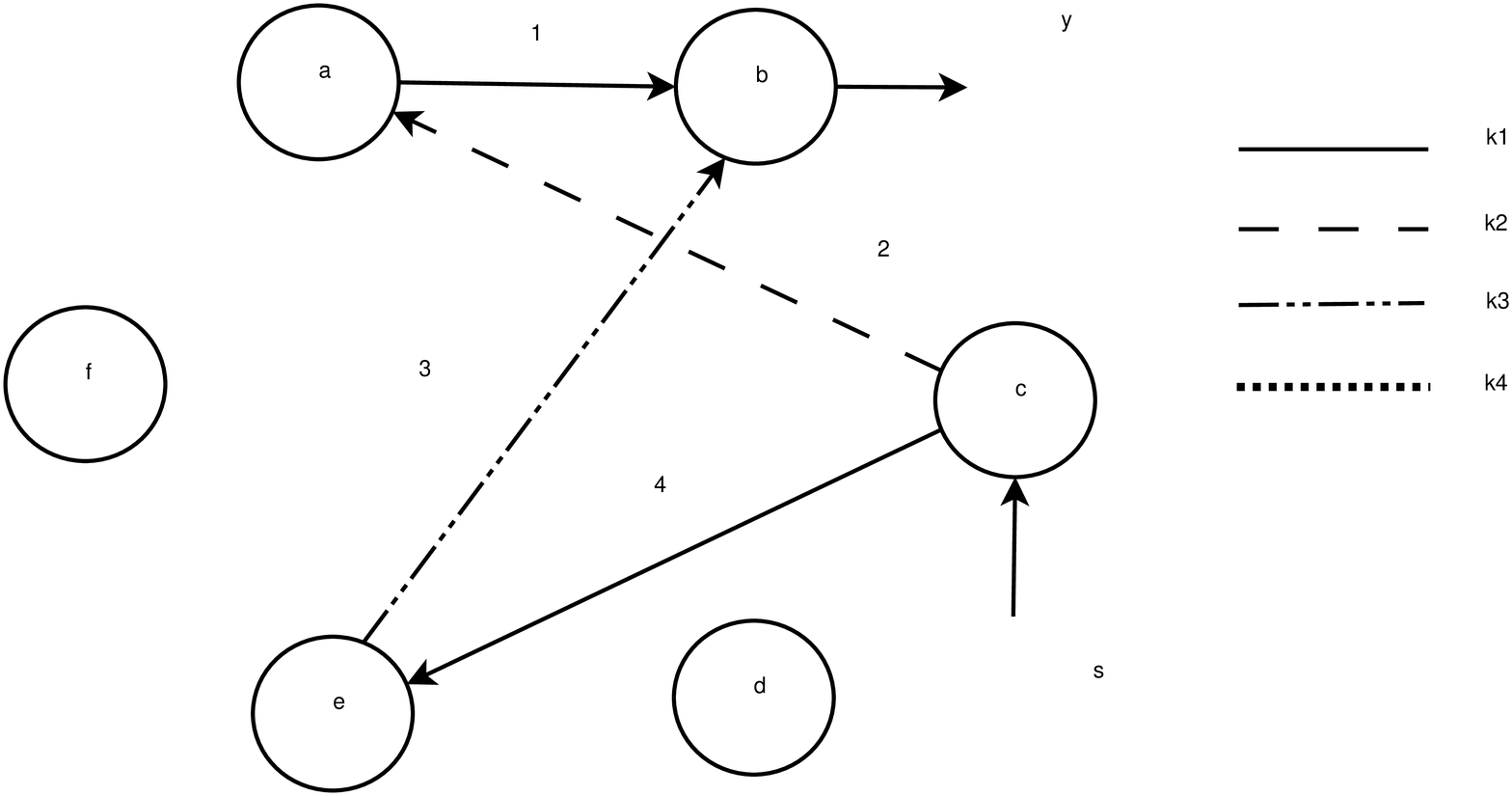}}
                \label{fig:sub_2}}
            \vspace{-0.2 cm}
      \caption{Data routes for \subref{fig:sub_1}  $d=1$, \subref{fig:sub_2} $d=2$ in Example~2.}
      \vspace{-0.2 cm}
     \label{fig:GP_D1}
\end{figure}

Setting the node power budgets to $P=20$~dBm and assuming that both rates have equal weights, $w_3^{(2)}=w_4^{(1)}=1$, the joint design algorithm in Section~\ref{se:RPC_w/o_time_GP_alg} yields a sum-rate of 9.1~b/s/Hz. The data routes and power allocations obtained by this algorithm are shown  in Figure~\ref{fig:GP_D1} and Table~\ref{tb:power}, respectively. For instance, in Figure~\ref{fig:GP_D1}, subcarrier $k=1$ is shown to be used twice and due to the half-duplex constraint, transmission and reception take place over distinct subcarriers at each node.
We will later show  the advantage of the proposed algorithm
over the algorithms in which frequency-reuse is not considered.
\hfill$\Box$

\begin{table}
\caption{Power Allocations (mW) in Example~2.} \label{tb:power}
\vspace{-.2 cm}
\begin{center}
\begin{tabular}{|l|l|l|l|}
\hline
$n=1$ &$n=3$ &$n=4$ &$n=5$ \\
	\hline
$q_{1,1}=12$   &  $q_{11,2}=45$ & $q_{16,4}=100$  & $q_{22,3}=25$ \\
               &  $q_{14,1}=55$ &                  &              \\
 \hline
\end{tabular}
\end{center}
\vspace{-.7 cm}
\end{table}

\textit{Example 3: (Average Weighted-Sum Rate Comparison)}
 \rvn{In this example, we
use Monte Carlo simulations to evaluate the average performance of the joint designs with and without time-sharing when the channels are time-varying rather than static as in Examples~1 and~2.}
We consider a network with $N=4$ nodes in which nodes 3 and 4 wish to communicate with nodes 2 and 1, respectively, over $K=4$ subcarriers. \edt{The number of simultaneous transmissions is set to $I=3$ and the} simulation results are averaged over 10 independent network realizations.

\begin{figure}        \centering
\psfrag{without time-sharing, with frequency reuse, proposed}[cl][cl][1.5]{w/o time-sharing, w. freq. reuse, proposed}
\psfrag{without time-sharing, without frequency-reuse, journal}[cl][cl][1.5]{w/o time-sharing, w/o freq. reuse, \cite{journal_1}}
\psfrag{with time-sharing, without frequency-reuse, journal}[cl][cl][1.5]{w. time-sharing, w/o freq. reuse, \cite{journal_1}}
\psfrag{with time-sharing,with frequency-reuse,proposed}[cl][cl][1.5]{w. time-sharing, w. freq. reuse, proposed}
\psfrag{a}[][][1.5]{$\begin{matrix}\\P~\text{[dBm]}\end{matrix}$ }
\psfrag{b}[][][1.5]{$\begin{matrix} \text{Weighted-sum rate (b/s/Hz)} \\ \\  \end{matrix}$ }
\psfrag{xx}[][][1.5]{$w_3^{(2)}=w_4^{(1)}$ }
\psfrag{zz}[][][1.5]{$w_3^{(2)}=5w_4^{(1)}$ }
\psfrag{yy}[][][1.5]{$w_4^{(1)}=5w_3^{(2)}$}
         \subfigure[]{
                \centering
                \resizebox{.45\textwidth}{!}{\includegraphics[width=\textwidth]{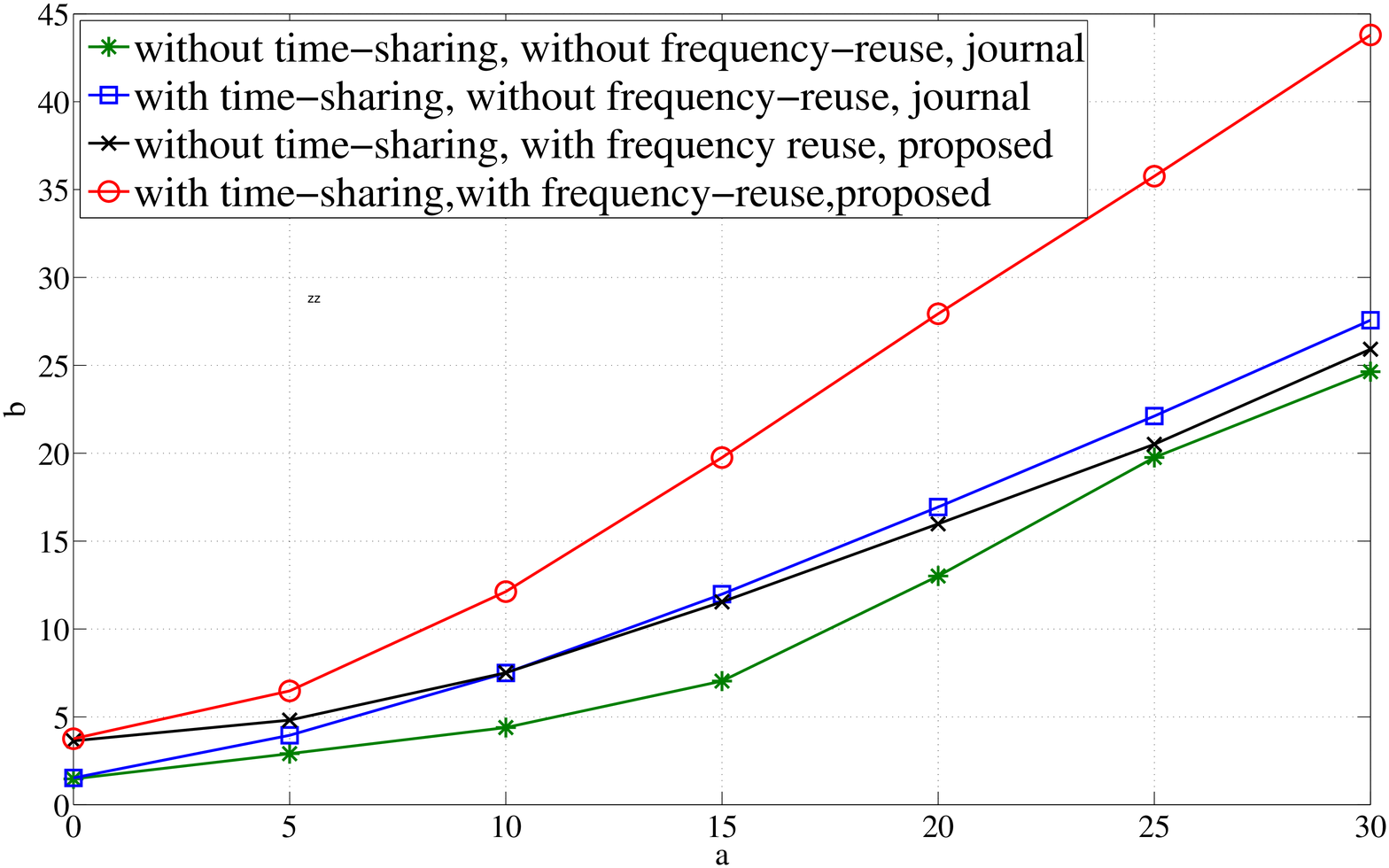}}
                \label{fig:w2}}
     \subfigure[]{
                \centering
                \resizebox{.45\textwidth}{!}{\includegraphics[width=\textwidth]{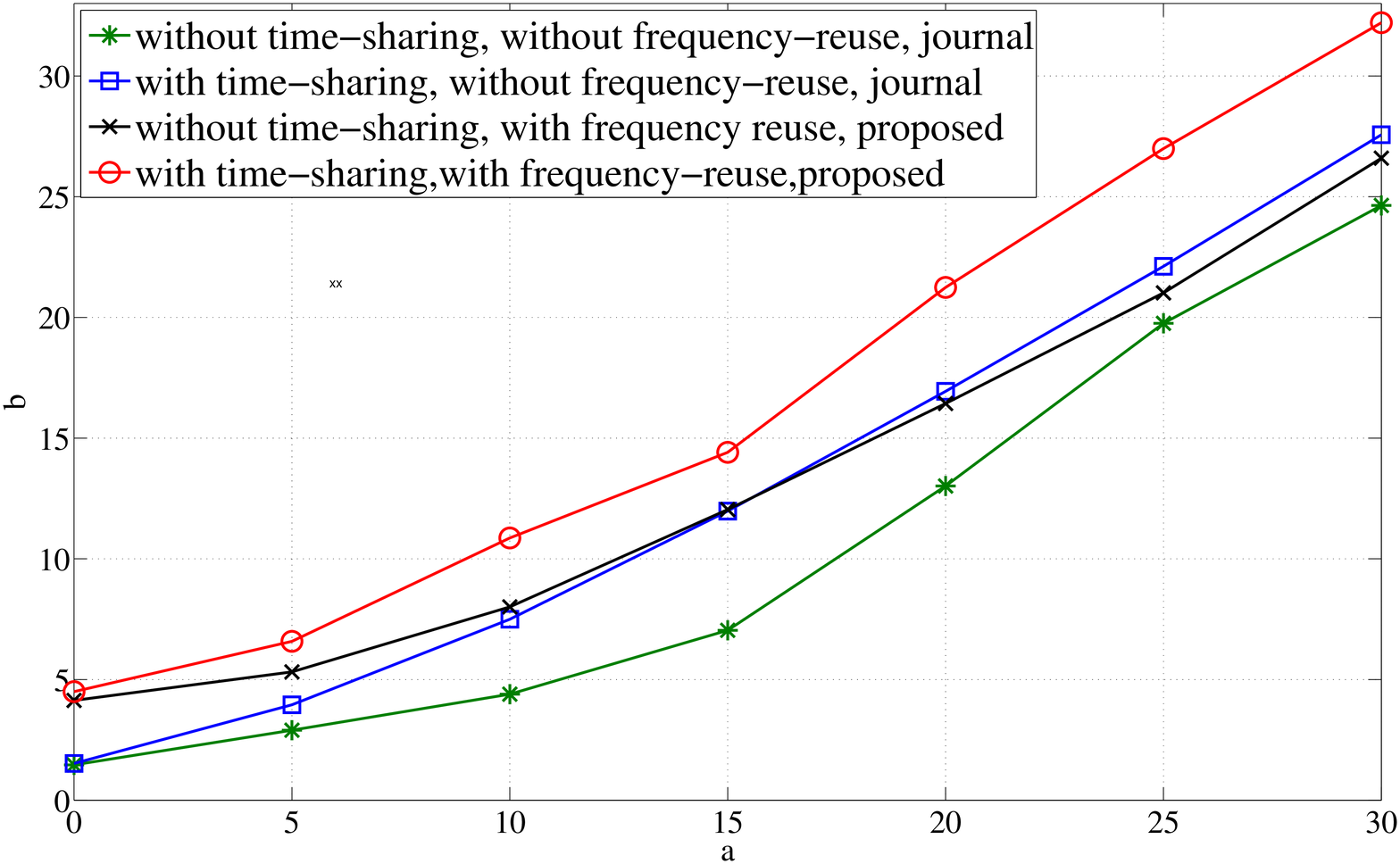}}
                \label{fig:w1}}
     \vspace{-0.2 cm}
      \caption{Average weighted-sum rate comparison for \subref{fig:w2} $w_3^{(2)}=5w_4^{(1)}$, and \subref{fig:w1}  $w_3^{(2)}=w_4^{(1)}$. } 
      \vspace{-.6 cm}
     \label{fig:graph}
\end{figure}

 The average weighted-sum rates yielded by the algorithms in Sections~\ref{se:RPC_w/o_time_GP_alg} and~\ref{se:RPC_w_time_GP_alg} for the values of $P$ ranging from 0 to 30 dBm are depicted in Figures~\ref{fig:w2} and~\ref{fig:w1} 
 for the cases of $w_3^{(2)}=5w_4^{(1)}$ and $w_3^{(2)}=w_4^{(1)}$, respectively. 
 These figures also provide a comparison
 with the weighted-sum rates yielded by the designs in which frequency-reuse is not considered~\cite{journal_1}.

 \new{As can be seen from Figure~\ref{fig:graph}, the weighted-sum rate yielded by the joint design with both time-sharing and frequency-reuse outperforms the  designs in which either time-sharing or frequency-reuse is exclusively considered, \rvn{but at the expense of increased complexity}. For instance, Figure~\ref{fig:w1} suggests that, at the sum-rate of 12~b/s/Hz, the proposed design with both time-sharing and frequency-reuse yields a power advantage of 4~dBm over the designs in which either time-sharing or frequency-reuse is exclusively considered and a power advantage of 8~dBm over the design in which neither of these techniques is considered. 
  This figure also suggests that, for values of $P$ less than 15 dBm, the design with frequency-reuse but without time-sharing yields better performance than the design with time-sharing but without frequency-reuse in~\cite{journal_1}. However, for values of $P$ higher than 15 dBm, the design with  time-sharing but without frequency-reuse performs better than the one with frequency-reuse but without time-sharing.
  This phenomenon can be attributed to the effect of interference. At low powers, the effect of interference is small and frequency-reuse  performs generally better than time-sharing. In contrast, at high powers, the effect of interference is more severe and time-sharing performs generally better than frequency-reuse.
  As expected,
  the design with neither time-sharing nor frequency-reuse has inferior performance.} 
\hfill$\Box$

\rvn{\textit{Example 4: (Joint Routing and Resource Allocation: Generalized Algorithm)}
In this example, we evaluate the performance of the algorithm developed in \edt{Section~\ref{se:RPC_w_time_GP_alg}}. 
We consider a snapshot of a network with $N =5$ nodes and $L=20$ links (links with a distance more than 150~m are neglected). In this network nodes 3 and 4 wish to communicate with nodes 2 and 1, respectively, over $K = 4$ subcarriers.
\edt{The number of simultaneous transmissions is set to $I=20$, $I=3$ and $I=2$, which results in $\Gamma$ with $2^{20}-1$,  190 and 110 variables, respectively.}

The sum-rate yielded by the \edt{generalized} algorithm with \edt{different values of $I$} is depicted in Figure~\ref{fig:modified}.
For comparison, this figure also shows the rates yielded by the
special case in Section~\ref{se:RPC_w/o_time}. As can be seen from Figure~\ref{fig:modified}, the algorithm \edt{with $I=2$ and 3} yields rates that are slightly less than the rate yielded by the algorithm \edt{with $I=L$}, however  with a significantly less computational complexity. In fact, the complexity of the algorithm \edt{with $I\ll L$ is polynomial, whereas that of the algorithm with $I=L$ is exponential in $L$.}
This feature renders the algorithm \edt{with $I\leq3$} more attractive for designing large networks with potentially rapid channel variations.
From Figure~\ref{fig:modified} it can be seen that the gap between the rates yielded by \edt{different values of $I$} decreases as the power budget increases. This is  because as power increases, interference becomes more severe, which causes the reuse of a particular subcarrier on multiple links less beneficial.
It can be also seen from this figure that, most of the gain of frequency-reuse is mustered by only considering two \edt{or three simultaneous transmissions, i.e., $I\leq3$}. \edt{This  implies that increasing $I$ trades complexity for performance. In particular, as $I$ increases, the performance of the algorithm becomes closer to that of the one with $I=L$, but at the expense of increased complexity.}

\begin{figure}[htb]\centering
\vspace{-.5 cm}
\psfrag{without frequency-reuse (I=0)}[cl][cl][3.3]{without freq. reuse $(I=1)$}
\psfrag{limited frequency-reuse (I=3)}[cl][cl][3.3]{limited freq. reuse $(I=3)$}
\psfrag{limited frequency-reuse (I=2)}[cl][cl][3.3]{limited freq. reuse $(I=2)$}
\psfrag{complete frequency-reuse (I=20)}[cl][cl][3.3]{complete freq. reuse $(I=20)$}
\psfrag{a}[][][3.5]{$\begin{matrix}\\P~\text{[dBm]}\end{matrix}$ }
\psfrag{b}[][][3.5]{$\begin{matrix} \text{Sum-rate (b/s/Hz)} \\ \\  \end{matrix}$ }
\resizebox{0.45\textwidth}{!}{\includegraphics*{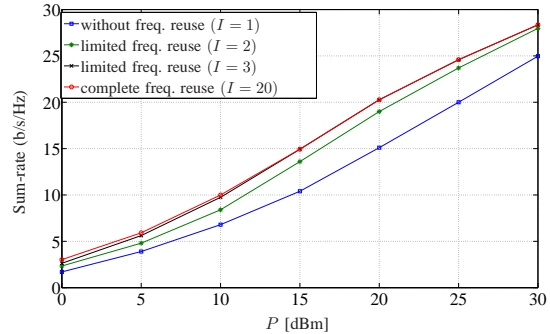}}
\caption{Sum-rate generated by the generalized algorithm for different values of $I$. }
\vspace{-.2 cm}
\label{fig:modified}
\end{figure}

 In Figure~\ref{fig:convergence} we investigate the convergence behaviour of the generalized algorithm. We consider an instance of a network in which  the power budget of all nodes is set to $P=10$~dBm.
It can be seen from this figure that, in addition to being significantly less computationally demanding, \edt{the algorithm with a lower value of $I$} exhibits considerably faster convergence than that of the one \edt{with a higher value of $I$.} This convergence can be further ameliorated by choosing the initial point more carefully, for instance, by choosing this point to be the solution yielded by algorithm in~\cite{journal_1} for the  case with no  frequency-reuse.

\edt{To illustrate the effect of random initialization of the generalized algorithm, in Figure~\ref{fig:initial} the value of the objective to which the generalized algorithm with $I=2$ converged is shown for  80 random instances of feasible initial points, $(q_{\ell k}^{(0)},\Gamma^{(0)})\in [0,P]^{LK}\times [0,1]^{|\Gamma|}$. 
 It can be seen from this figure that
 although the algorithm is relatively sensitive to the choice of the initial point, finding initial points that result in `good' local maxima is generally easy.
}
\hfill$\Box$

\begin{figure}[htb]\centering
\vspace{-.5 cm}
\psfrag{final value (I=20)}[cl][cl][3.3]{final value (I=20)}
\psfrag{final value (I=2)}[cl][cl][3.3]{final value (I=2)}
\psfrag{each iteration (I=2)}[cl][cl][3.3]{each iteration (I=2)}
\psfrag{each iteration (I=20)}[cl][cl][3.3]{each iteration (I=20)}
\psfrag{experiment}[cl][cl][3.3]{experiment}
\psfrag{top 35}[cl][cl][3.3]{Top $10\%$}
\psfrag{b}[][][3.5]{$\begin{matrix} \text{Sum-rate (b/s/Hz)} \\ \\  \end{matrix}$ }
\subfigure[]{
                \centering
                \psfrag{a}[][][3.5]{Number of iterations }
                \resizebox{0.45\textwidth}{!}{\includegraphics*{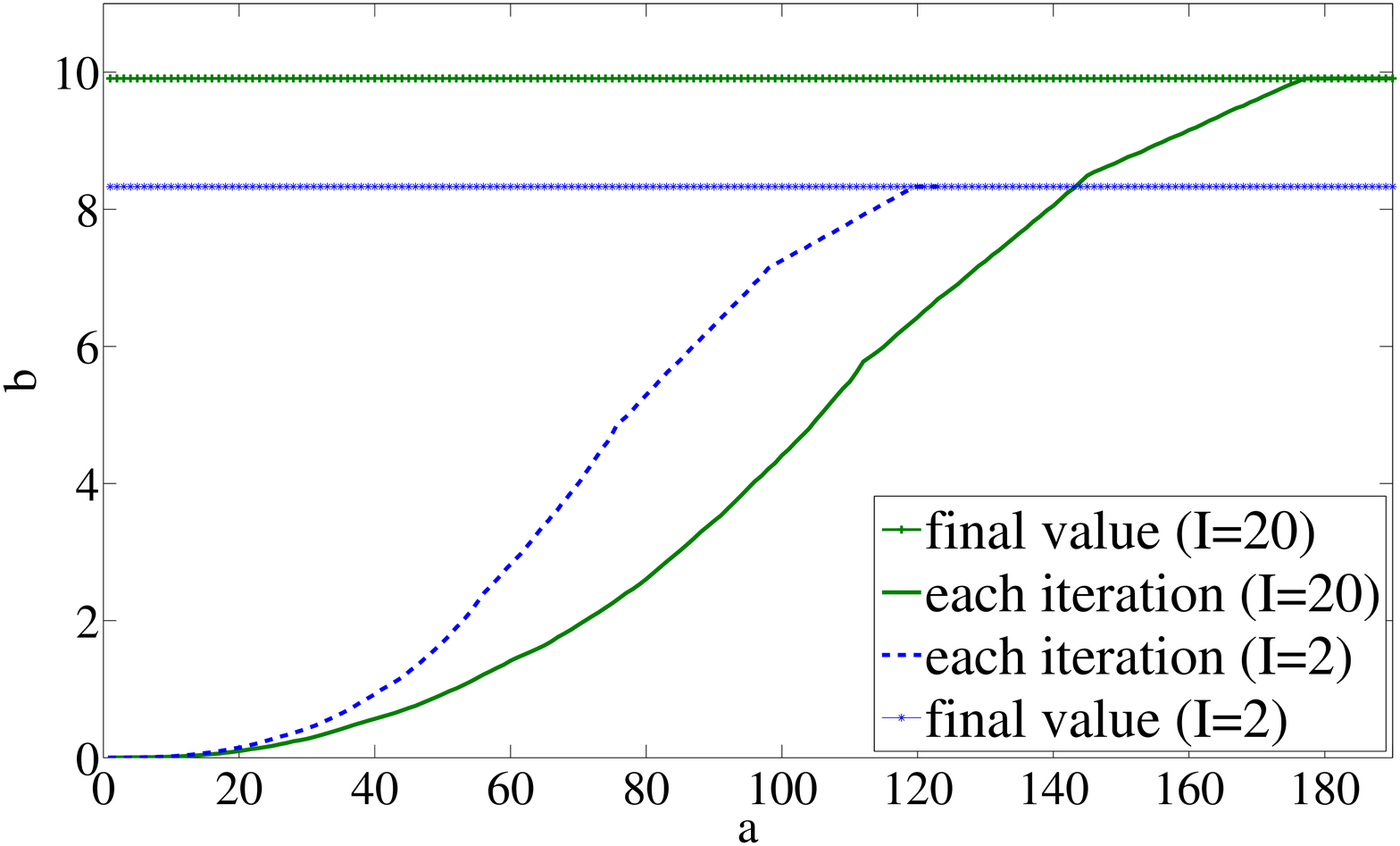}}
                \label{fig:convergence}}
\subfigure[]{
                \centering
                \psfrag{a}[][][3.5]{Number of experiments }
\resizebox{0.45\textwidth}{!}{\includegraphics*{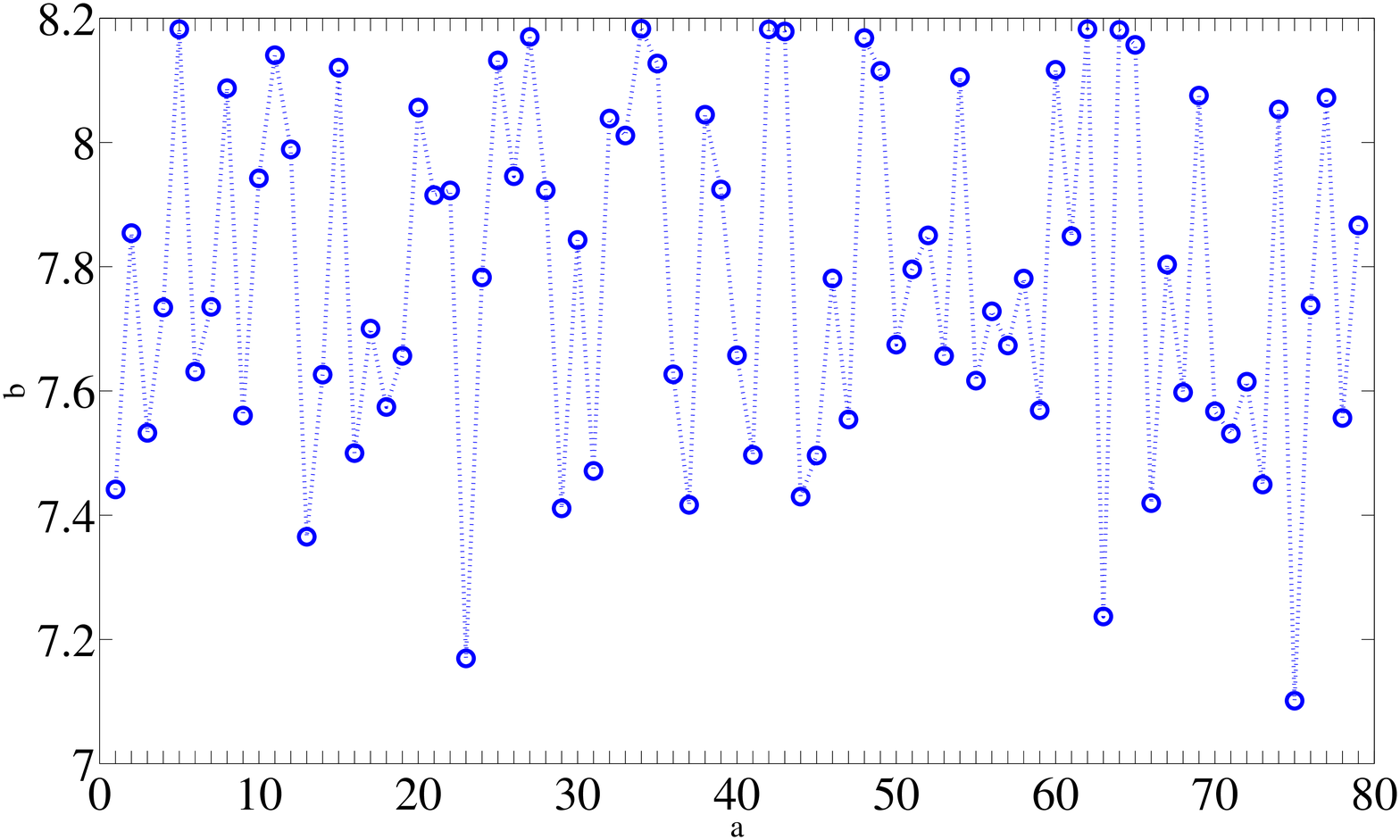}}
\label{fig:initial}}
\vspace{-.2 cm}
\caption{\subref{fig:convergence} Convergence behaviour and \subref{fig:initial} performance of the generalized algorithm with different initial points. }
\vspace{-.2 cm}
\label{fig:general}
\end{figure}

}

\begin{figure}[htb]\centering
\vspace{-.2 cm}
\psfrag{without time-sharing, with frequency reuse, proposed}[cl][cl][2.6]{w/o time-sharing, w. freq. reuse}
\psfrag{without time-sharing, without frequency-reuse, journal}[cl][cl][2.6]{w/o time-sharing, w/o freq. reuse, \cite{journal_1}}
\psfrag{with time-sharing, without frequency-reuse, journal}[cl][cl][2.6]{w. time-sharing, w/o freq. reuse, \cite{journal_1}}
\psfrag{with time-sharing,with frequency-reuse,proposed}[cl][cl][2.6]{w. time-sharing, w. freq. reuse, $I=20$}
\psfrag{I=2}[cl][cl][2.6]{w. time-sharing, w. freq. reuse, $I=2$}
\psfrag{I=3}[cl][cl][2.6]{w. time-sharing, w. freq. reuse, $I=3$}
\psfrag{a}[][][3.5]{$\begin{matrix}\\s_4^{(1)}~\text{(b/s/Hz)}\end{matrix}$ }
\psfrag{b}[][][3.5]{$\begin{matrix} s_3^{(2)}~\text{(b/s/Hz)}\\ \\ \end{matrix}$ }
\resizebox{0.45\textwidth}{!}{\includegraphics*{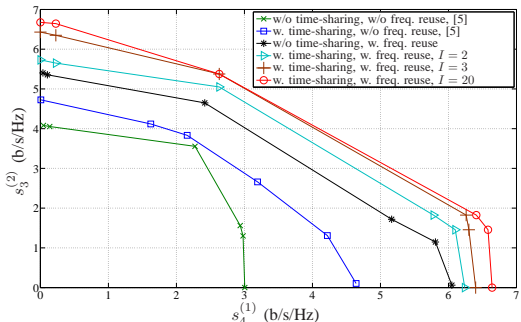}}
\vspace{-.2 cm}
\caption{Rate-region comparison. }
\vspace{-.4 cm}
\label{fig:GP_cap}
\end{figure}

\textit{Example 5: (Average Rate-Region Comparison)}
 In this example we provide the rate regions that can be achieved by the algorithms in Sections~\ref{se:RPC_w/o_time_GP_alg} and~\ref{se:RPC_w_time_GP_alg}, when $P=10$ dBm. These regions are obtained by varying the weights $(s_4^{(1)},s_3^{(2)})$ over the unit simplex, i.e., $\bigl\{(w_4^{(1)},w_3^{(2)})|w_4^{(1)}\geq 0,\, w_3^{(2)}\geq0,\, w_4^{(1)}+w_3^{(2)}=1\bigr\}$, and
  are depicted in Figure~\ref{fig:GP_cap}.
  \new{A comparison between these rate regions and the ones corresponding to the case when frequency-reuse is not considered~\cite{journal_1} is also provided in this figure. 
 As can be seen from Figure~\ref{fig:GP_cap}, the rate region corresponding to the design with both time-sharing and frequency-reuse properly contains the rate regions corresponding to the designs in which either time-sharing or frequency-reuse is exclusively used.} 
\edt{It can be also seen that restricting the number of simultaneous transmissions to be less than three suffices to achieve most of the  frequency-reuse gain and with less computational complexity.}
 %
 \hfill$\Box$

\section{Conclusion}  \label{se:con}
\new{In this paper we focused on the joint optimization of data routes, subcarrier schedules and power allocation in a half-duplex multicarrier network when each subcarrier can be reused by multiple links. The goal is to maximize a weighted-sum of the rates communicated over the network. The considered network is generic in the  sense that it subsumes many structures including cellular and device-to-device communications as special cases. 
\rvn{We considered two instances of this problem: 1) when each subcarrier can be time-shared by multiple links; and 2) when time-sharing is not allowed and a subcarrier, once assigned to a set of links, will be used by those  links throughout the signalling interval.}
The joint design in the first instance results in superior performance but with high complexity.
The second instance is a special case of the first one and can be parameterized using a significantly smaller number of variables. 

The joint design problem in both instances is nonconvex and locally optimal solutions are obtained using a GP-based monomial approximation technique. 
 Numerical results show that the designs developed in both instances yield performance that is significantly better than  that of their counterparts in which frequency-reuse is not allowed.}

\begin{appendices}
\section{The GP Standard Form and Monomial Approximation} \label{se:GP_mon_def}
\subsubsection{The GP Standard Form}   \label{se:GP_def}
For self-containment, in this appendix we will review the standard GP form. A GP optimization problem can be readily transformed to an efficiently solvable convex one. To provide the standard form of a GP,
let $z\in\mathbb{R}^n$ be  a vector of positive entries.  A monomial in $z$ is defined to be a function of the form $c_0\prod_{i}z_i^{\alpha_i}$ and a posynomial in $z$ is defined to be a function of the form $\sum_{j=1}^J{c_j}\prod_{i=1}^nz_i^{\alpha_{ij}}$, where $c_j>0$, $\{\alpha_i\}$ and $\{\alpha_{ij}\}$,   are arbitrary constants, $j=0,1,\ldots,J$, and $i=1,\ldots,n$. A standard GP~\cite{CTPOJ07,Boyd,BKVH07} is an optimization of the form:
\begin{align}
\min_z\;\; &f_0(z),\nonumber\\
\text{subject to} \;\; &f_i(z)\leq 1, \qquad\qquad i=1,\ldots,m,\label{eq: std-GP}\\
&g_i(z)=1, \qquad\qquad i=1,\ldots,p,\nonumber
\end{align}
where $\{f_i\}$  are posynomials and $\{g_i\}$ are monomials. 

\subsubsection{Monomial Approximation} \label{se:mon_def}
A monomial approximation of a differentiable function $h(z)\geq 0$ near $z^{(0)}$ is given by its first order Taylor expansion in the logarithmic domain~\cite{CTPOJ07,BKVH07}. Defining \(\beta_i=\frac{z_i^{(0)}}{h(z^{(0)})}\frac{\partial h}{\partial z_i}\big|_{z=z^{(0)}}\), we have
$
\mathcal{M}\left(h(z)\right) = h(z^{(0)})\prod_{i=1}^n\Bigl(\frac{z_i}{z_i^{(0)}} \Bigr)^{\beta_i}, 
$
 where $\mathcal{M}(\cdot)$ is the monomial approximation. This approximation will be used to provide local GP approximations in the neighbourhood of a given initial point. 

\section{Proof of Proposition~\ref{pr:GP}} \label{se:slf-crdn}
\rvn{For simplicity, we will begin by proving the second statement of Proposition~\ref{pr:GP}.} 
In this proof, we will show that, by including redundant constraints, the log-barrier function of the problems in~\eqref{eq: JRPA-GP} and~\eqref{eq:GP} can be cast in a self-concordant form, which has the following definition~\cite{Boyd}: 
\begin{definition} \label{def:slf-coc}
A function $f:\mathbb{R}^n\to \mathbb{R}$ is said to be self-concordant if, for all $x, v\in\mathbb{R}^n$, $s\in\mathbb{R}$ such that $x+sv$ is in the domain of $f$ and $\Bigl|\frac{\partial^{3}}{\partial s^3} f(x+sv)\Bigr| \leq 2 \frac{\partial^{2}}{\partial s^2} f(x+sv)^{3/2}$.
\hfill$\Box$
\end{definition}

\subsection{Proof of the Second Statement of Proposition~\ref{pr:GP}} \label{se:pr1-proof}
To determine the complexity of solving the problem in~\eqref{eq: JRPA-GP}, we begin by converting this problem into a convex one. Using standard exponential transformations, we write
\begin{alignat}{2} \label{eq:exp}
t_n^{(d)}&=\exp \left(\ln(2)s_n^{(d)}\right),\qquad & n\in\mathcal{N}\setminus\{d\},\, d\in\mathcal{D},\nonumber\\
r_{\ell k}^{(d)}&=\exp \Bigl( \ln(2)\frac{x_{\ell k}^{(d)}}{W}\Bigr), \qquad & \ell\in\mathcal{L},\, k\in\mathcal{K},\, d\in\mathcal{D}, \nonumber \\
y_{\ell k}&=\exp(q_{\ell k}), & \ell\in\mathcal{L},\, k\in\mathcal{K}.
\end{alignat}
Substituting the variables in~\eqref{eq: JRPA-GP} with the ones in~\eqref{eq:exp} and taking the logarithm of the obtained objective and constraints result in a convex optimization which can be solved efficiently using the IPM technique. To use this technique, a log-barrier function is synthesized from  the objective and inequality constraints. The complexity analysis of the IPM technique is simplified when the log-barrier function is self-concordant~\cite{Boyd}, cf., Definition~\ref{def:slf-coc}.
The log-barrier function corresponding to the convex form of~\eqref{eq: JRPA-GP} can be written as
\begin{equation}
\phi=-t\sum_{n}\sum_{d} w_n^{(d)}s_n^{(d)} + \psi, \label{eq:lg-br}
\end{equation}
where $\psi$ represents the component of the log-barrier function associated with the inequality constraints in the convex form of~\eqref{eq: JRPA-GP}. To examine whether $\phi$ is self-concordant, we note that the converted objective
and the inequality constraints corresponding to~\eqref{eq:non-neg-flow-gp},~\eqref{eq:non-neg-rate-gp} and the relaxed versions of~\eqref{eq:non-mulcast_1} and~\eqref{eq:non-mulcast} are linear and therefore their corresponding components in the log-barrier function are self-concordant~\cite{Boyd}. Hence it remains to consider the self-concordance for the constraints in~\eqref{eq:gp_pos} and~\eqref{eq:pow-bdg}. For simplicity, we write the posynomial constraint in~\eqref{eq:gp_pos} in the standard form in~\eqref{eq: std-GP}. 
After changing the variables and taking the logarithm  of both sides, this constraint can be written in a general form as
\begin{equation}
\log \bigl(\sum_i \exp(a_i\alpha_i +b_i\beta_i  + c_i) \bigr) \leq 0,  \label{eq:gp_cap_std}
\end{equation}
where $\{\alpha_i\},\{\beta_i\}$ are the optimization variables and $\{a_i\},\{b_i\},\{c_i\}$ are constants.
The component corresponding to the constraint in~\eqref{eq:gp_cap_std} in the log-barrier function can now be expressed as
\begin{equation}
- \log \bigl( -\log \sum_i \exp(a_i\alpha_i +b_i\beta_i  + c_i) \bigr).  \label{eq:gp_lg_br}
\end{equation}
To ensure that~\eqref{eq:gp_lg_br} is self-concordant, 
we introduce auxiliary variables, $\lambda_i$, to bound the exponentially transformed variables in~\eqref{eq:gp_cap_std}. Using these new variables, the constraint in~\eqref{eq:gp_cap_std} can be replaced with the following set of constraints~\cite{Boyd}:
\begin{align} \label{eq:aux_cons}
& \sum_i \lambda_i \leq 1, \nonumber \\
& \lambda_i \geq 0, \nonumber \\
& a_i\alpha_i+b_i\beta_i  + c_i -\log \lambda_i \leq 0.
\end{align}
Now the associated log-barrier function of the constraints in~\eqref{eq:aux_cons} can be shown to be self-concordant, cf.,~\cite[Example 9.8]{Boyd}.
For the constraints in~\eqref{eq:pow-bdg}, we follow the steps analogous to the ones used with the constraints in~\eqref{eq:gp_pos}. In particular, by introducing new auxiliary variables, we construct a self-concordant log-barrier function. Using this function, the complexity can be shown to be proportional to $m^{3.5}$, where $m$ is the number of inequality constraints. Hence, the complexity of solving~\eqref{eq: JRPA-GP} can be bounded by $\mathcal{O}\left(\left(LK(3L+2)+N+D(N-1)\right)^{3.5}\right)$, which completes the proof of the second statement of Proposition~\ref{pr:GP}.

\subsection{Proof of the First Statement of Proposition~\ref{pr:GP}} \label{se:pr2-proof}
The proof of the first statement of Proposition~\ref{pr:GP} follows from arguments similar to the one used in the proof of the second statement and is omitted for brevity.
\rvn{For the first statement,  the number of inequality constraints can be readily verified to be
$2LKN+N+D(N-1)+2K\sum_{i=1}^{I}{L\choose i}$,  which yields the first statement of Proposition~\ref{pr:GP}.}
\end{appendices}

\bibliographystyle{ieeetr}

\end{document}